\documentclass[usenatbib]{mnras}

\usepackage{newtxtext,newtxmath}

\usepackage[T1]{fontenc}
\usepackage{ae,aecompl}

\usepackage{graphicx}	
\usepackage{amsmath}	
\usepackage[usenames, dvipsnames]{color}
\usepackage{url}

\title[Massive Quiescent Galaxies]{Investigating The Growing Population of Massive Quiescent Galaxies at Cosmic Noon}

\author[Sherman et al.]{Sydney Sherman$^{1}$\thanks{E-mail: \texttt{ssherman@astro.as.utexas.edu}},
Shardha Jogee$^{1}$,
Jonathan Florez$^{1}$,
Matthew L. Stevans$^{1}$, \newauthor
Lalitwadee Kawinwanichakij$^{2,3,4,5}$, 
Isak Wold$^{6}$,
Steven L. Finkelstein$^{1}$, 
Casey Papovich$^{3,4}$, \newauthor
Robin Ciardullo$^{7,8}$,
Caryl Gronwall$^{7,8}$, 
Sof{\'i}a A. Cora$^{9,10}$, 
Tom{\'a}s Hough$^{9,10}$, and \newauthor
Cristian A. Vega-Mart{\'i}nez$^{11,12}$
\\
$^{1}$Department of Astronomy, The University of Texas at Austin, Austin, TX 78712 \\
$^{2}$Kavli Institute for the Physics and Mathematics of the Universe, The University of Tokyo, Kashiwa, Japan 277-8583 (Kavli IPMU, WPI)\\
$^{3}$Department of Physics and Astronomy, Texas A\&M University, College Station, TX, 77843-4242\\
$^{4}$George P. and Cynthia Woods Mitchell Institute for Fundamental Physics and Astronomy, Texas A\&M University, College Station, TX 77843\\
$^{5}$LSSTC Data Science Fellow\\
$^{6}$NASA Goddard Space Flight Center, Greenbelt, MD 20771\\
$^{7}$Department of Astronomy and Astrophysics, The Pennsylvania State University, University Park, PA 16802\\
$^{8}$Institute for Gravitation and the Cosmos, The Pennsylvania State University, University Park, PA 16802\\
$^{9}$Instituto de Astrof\'isica de La Plata (CCT La Plata, CONICET, UNLP), Observatorio Astron\'omico,\\
Paseo del Bosque, B1900FWA La Plata, Argentina\\
$^{10}$Facultad de Ciencias Astron\'omicas y Geof\'isicas, Universidad Nacional de La Plata, Observatorio Astron\'omico,\\
Paseo del Bosque, B1900FWA La Plata, Argentina\\
$^{11}$Instituto de Investigaci\'on Multidisciplinar en Ciencia y Tecnolog\'ia, Universidad de La Serena, Ra\'ul Bitr\'an 1305, La Serena, Chile\\
$^{12}$Departamento de F\'isica y Astronom\'ia, Universidad de La Serena, Av. Juan Cisternas 1200 Norte, La Serena, Chile
}

\date{Accepted 2020 October 9. Received 2020 September 30; in original form 2020 June 24}

\pubyear{2020}

\begin{document}
\label{firstpage}
\pagerange{\pageref{firstpage}--\pageref{lastpage}}
\maketitle

\begin{abstract}
We explore the buildup of quiescent galaxies using a sample of 28,469 massive ($M_\star \ge 10^{11}$M$_\odot$) galaxies at redshifts $1.5<z<3.0$, drawn from a 17.5 deg$^2$ area (0.33 Gpc$^3$ comoving volume at these redshifts). This allows for a robust study of the quiescent fraction as a function of mass at $1.5<z<3.0$ with a sample $\sim$40 times larger at log($M_{\star}$/$\rm M_{\odot}$)$\ge11.5$ than previous studies. We derive the quiescent fraction using three methods: specific star-formation rate, distance from the main sequence, and UVJ color-color selection. All three methods give similar values at $1.5<z<2.0$, however the results differ by up to a factor of two at $2.0<z<3.0$. At redshifts $1.5 < z < 3.0$ the quiescent fraction increases as a function of stellar mass. By $z=2$, only 3.3 Gyr after the Big Bang, the universe has quenched $\sim$25\% of $M_\star = 10^{11}$M$_\odot$ galaxies and $\sim$45\% of $M_\star = 10^{12}$M$_\odot$ galaxies. We discuss physical mechanisms across a range of epochs and environments that could explain our results. We compare our results with predictions from hydrodynamical simulations SIMBA and IllustrisTNG and semi-analytic models (SAMs) SAG, SAGE, and Galacticus. The quiescent fraction from IllustrisTNG is higher than our empirical result by a factor of $2-5$, while those from SIMBA and the three SAMs are lower by a factor of $1.5-10$ at $1.5<z<3.0$. 
\end{abstract}



\section{Introduction}
Understanding how galaxies are transformed from star-forming to quiescent is a key question in the study of galaxy evolution. The population of massive galaxies, with stellar masses $M_\star \ge 10^{11}$M$_\odot$, is thought to grow rapidly at early times (e.g., \citealt{Cowie1996}, \citealt{Bundy2006}) compared to less massive galaxies. Theoretical models have shown the important role that feedback plays in galaxy evolution (e.g., \citealt{Somerville1999}, \citealt{Cole2000}, \citealt{Bower2006}, \citealt{Croton2006}, \citealt{Somerville2008}, \citealt{Benson2012}, \citealt{Somerville2015} and references therein, \citealt{Croton2016}, \citealt{Naab2017} and references therein, \citealt{Weinberger2017}, \citealt{Cora2018}, \citealt{Knebe2018}, \citealt{Behroozi2019}, \citealt{Cora2019}, \citealt{Dave2019}), and without these feedback mechanisms, galaxies at present day are too massive compared to the observed galaxy population (e.g., \citealt{Weinberger2017}, \citealt{Pillepich2018b}). Compiling statistically significant samples of these massive galaxies at cosmic noon ($1.5 < z < 3.0$), a time when galaxy assembly progressed rapidly, can help to uncover the physical processes, specifically as a function of mass, driving the shift from a predominantly star-forming massive galaxy population to one that is primarily quiescent. 

The number and properties of massive quiescent galaxies in place by $z\sim2$ (only 3.3 Gyr after the Big Bang) provide important tests of galaxy evolution models. Theoretical models need to implement physical processes that can reproduce the rapid formation and early quenching of massive galaxies in such a short time following the Big Bang, along with the wide range of sizes, structures, and specific star-formation rates (sSFR; $\rm sSFR = \rm SFR/\rm M_\star$) (e.g., \citealt{Conselice2011}, \citealt{vanderWel2011}. \citealt{Weinzirl2011}, \citealt{vanDokkum2015}) seen in observational studies by $z\sim2$. Additionally, these models must simultaneously match the much slower growth of less massive systems. Cosmic noon ($1.5 < z < 3.0$) is a particularly important epoch to study galaxy evolution as this is a time when proto-clusters begin collapsing into the galaxy clusters seen at present day and the cosmic star-formation rate density and black hole accretion rate peak (e.g., \citealt{MadauDickinson2014}). 

In the extragalactic community galaxies are typically defined to be ``quiescent" if they have sufficiently low specific star-formation rate (e.g., \citealt{Fontanot2009}), lie a given distance below the main sequence (e.g., \citealt{Fang2018}, \citealt{Donnari2019}), or fall in a particular region of rest-frame color-color space (e.g., \citealt{Labbe2005}, \citealt{Wuyts2007}, \citealt{Williams2009}, \citealt{Muzzin2013}). It should be noted that these definitions of quiescence do not necessarily imply an abrupt cessation of star-formation. In the literature, a wide variety of mechanisms have been invoked for quenching star-formation in galaxies (e.g., \citealt{Kawinwanichakij2017}, \citealt{Man2018}, \citealt{Papovich2018}), and they fall into two inter-related categories. The first involves processes that \textit{accelerate} star-formation and the consumption of gas (e.g., \citealt{Man2018}) by driving gas to the circumnuclear regions where the gas reaches high densities and fuels rapid star-formation. These processes include major and minor mergers (e.g., \citealt{Mihos1994}, \citealt{Mihos1996}, \citealt{Jogee2009}, \citealt{Robaina2010}, \citealt{Hopkins2013}), tidal interactions (e.g., \citealt{Barnes1992} and references therein, \citealt{Gnedin2003}), and spontaneously or tidally induced bars (e.g., \citealt{Sakamoto1999}, \citealt{Jogee2005}, \citealt{Peschken2019}). The second category involves processes that \textit{suppress} star-formation. This includes ram pressure stripping (e.g., \citealt{Gunn1972}, \citealt{Giovanelli1983}, \citealt{Cayatte1990}, \citealt{Koopmann2004}, \citealt{Crowl2005}, \citealt{Singh2019}) where the intra-cluster medium removes cold gas from a galaxy traveling within a cluster, tidal stripping (\citealt{Moore1996}, \citealt{Moore1998}), or starvation and strangulation \citep{Larson1980}. This second category also includes mechanisms such as stellar (e.g., \citealt{Ceverino2009}, \citealt{Vogelsberger2013}, \citealt{Hopkins2016}, \citealt{Nunez2017}) and AGN feedback (e.g., \citealt{Hambrick2011}, \citealt{Fabian2012}, \citealt{Vogelsberger2013}, \citealt{Choi2015}, \citealt{Hopkins2016}) that heat, redistribute, and/or expel gas residing in a galaxy or gas accreting onto the galaxy.

Previous observational studies (e.g., \citealt{Kriek2006}, \citealt{Muzzin2013}, \citealt{Martis2016}, \citealt{Tomczak2016}) have typically focused on the quiescent fraction as a function of redshift for large mass bins and have found that the quiescent fraction at the massive end increases towards present day. A significant challenge faced by these studies is that their small area provides small samples of massive ($M_\star \ge 10^{11}$M$_\odot$) galaxies, which are rare in number density. Additionally, small area studies suffer from the effects of cosmic variance (e.g., \citealt{Driver2010}, \citealt{Moster2011}), which, in combination with small sample sizes leads to large errors in measures of the quiescent fraction of massive galaxies. Without many galaxies at the high mass end, large mass bins are typically used which leads to broad conclusions about the nature of the population of quenched massive galaxies. 

In this work, we present the quiescent fraction of massive galaxies measured three ways (specific star-formation rate, distance from the main sequence, and UVJ color-color selection) using a sample of 28,469 massive ($M_\star \ge 10^{11}$M$_\odot$) galaxies at $1.5 < z < 3.0$ which are uniformly selected from a 17.5 deg$^2$ area which probes a comoving volume of $\sim$0.33 Gpc$^3$ at these redshifts. With this large sample of star-forming and quiescent galaxies selected over a large comoving volume we are able to significantly reduce the error due to Poisson statistics and cosmic variance. We are uniquely suited to split our results at the high mass end as a function of mass, which allows us to place constraints on the quenching mechanisms at play across a range of galaxy stellar masses. This is a significant improvement over previous works that used large mass bins (typically all galaxies with $M_\star \ge 10^{11}$M$_\odot$ in a single bin) due to small sample sizes. Our $1.5 < z < 3.0$ sample is a factor of $\sim$40 larger at log($M_\star$/M$_\odot$) $\ge$ 11.5 than samples from previous studies. We discuss the physical processes which may contribute to the trends found in our empirical quiescent fraction as a function of mass.

We also compare our empirical quiescent fraction with predictions from two types of theoretical models: hydrodynamical models from IllustrisTNG (\citealt{Pillepich2018b}, \citealt{Springel2018}, \citealt{Nelson2018}, \citealt{Naiman2018}, \citealt{Marinacci2018}) and SIMBA \citep{Dave2019}, and semi-analytic models (SAMs) SAG \citep{Cora2018}, SAGE \citep{Croton2016}, and Galacticus \citep{Benson2012}. To further constrain models of galaxy evolution, we also compute the galaxy stellar mass function for the total sample of galaxies, as well as the star-forming and quiescent galaxy populations. Comparisons of our empirical quiescent fraction and stellar mass function with predictions from theoretical models provides powerful constraints on the implementation of baryonic physics in these models (particularly star-formation and feedback models).

This paper is organized as follows. In Section \ref{sec:data} we introduce the data used in this work and in Section \ref{sec:analysis} we detail the data analysis and SED fitting procedure. Our empirical results are presented in Section \ref{sec:emperical_results}, including the quiescent fraction as a function of mass (Section \ref{sec:qf_shela}) and redshift (Section \ref{sec:qf_shela_vz}), and the empirical galaxy stellar mass function (Section \ref{sec:shela_smf}). In Section \ref{sec:discussion} we discuss the physical mechanisms that may quench the galaxies in our sample (Section \ref{sec:quench_mech}). In Sections \ref{sec:qf_mass_vTheory}, \ref{sec:qf_z_vTheory}, and \ref{sec:theory_smf} we compare our empirical results on the quiescent fraction as a function of mass and redshift, and the galaxy stellar mass function with predictions from theoretical models. Finally, we summarize our results in Section \ref{sec:summary}. Throughout this work we adopt a flat $\Lambda$CDM cosmology with $h = 0.7$, $\Omega_{\rm m} = 0.3$, and $\Omega_{\Lambda} = 0.7$. 

\section{Data}
\label{sec:data}
The data used in this work come from large area surveys, covering $\sim$17.5 deg$^2$, in the \textit{Spitzer}-HETDEX Exploratory Large Area (SHELA;  \citealt{Papovich2016}, \citealt{Wold2019}) footprint. The five primary photometric data sets used in this study come from the Dark Energy Camera (DECam) \textit{u,g,r,i,z} \citep{Wold2019}, NEWFIRM $K_s$ (PI Finkelstein; Stevans et al.\ submitted), \textit{Spitzer}-IRAC 3.6 and 4.5$\mu$m (PI Papovich; \citealt{Papovich2016}), \textit{Herschel}-SPIRE (HerS, \citealt{Viero2014}) far-IR/submillimeter, and XMM-Newton and Chandra X-ray Observatory X-ray data from the Stripe 82X survey (\citealt{LaMassa2013a}, \citealt{LaMassa2013b}, \citealt{Ananna2017}, the X-ray data cover $\sim$11.2 deg$^2$). We gain additional photometric coverage in the near-IR with $J$ and $K_s$ data from the VICS82 Survey \citep{Geach2017}. Optical spectroscopy in this region is being acquired by the Hobby Eberly Telescope Dark Energy Experiment (HETDEX, \citealt{Hill2008}). 

In this work we utilize data from DECam \textit{u,g,r,i,z}, NEWFIRM $K_s$, VICS82 $J$ and $K_s$, and \textit{Spitzer}-IRAC 3.6 and 4.5$\mu$m to perform spectral energy distribution (SED) fitting of our sample (see Section \ref{sec:analysis}). Our catalog (Stevans et al. Submitted) is $K_s$-selected by implementing Source Extractor (SExtractor; \citealt{BertinArnouts1996}) on the NEWFIRM $K_s$ imaging, which reach a 5$\sigma$ depth of 22.4 AB mag. With object locations determined from running SExtractor on the NEWFIRM $K_s$ imaging, forced photometry is performed on the DECam \textit{u,g,r,i,z} (r-band 5$\sigma$ depth is r = 24.5 AB mag; \cite{Wold2019}) and \textit{Spitzer}-IRAC 3.6 and 4.5$\mu$m imaging (with $5\sigma$ depth of 22 AB mag in both filters; \cite{Papovich2016}) using the Tractor \citep{tractor}. For a complete and detailed description of catalog construction, see Stevans et al. (submitted) and \cite{Kawinwanichakij2020}, as well as \cite{Wold2019} who implement a similar procedure. 

We note that the sample used in this paper is constructed in a similar way to that from \cite{Sherman2020}, however one key difference has allowed for the study contained in this work. \cite{Sherman2020} performed a study of massive ($M_\star \ge 10^{11}$M$_\odot$) star-forming galaxies in the same footprint used here, but their analysis was limited to star-forming galaxies, as the catalog used \citep{Wold2019} was \textit{riz}-selected. Since that publication, NEWFIRM $K_s$ data have become available in the footrprint, which allows for the selection of both the star-forming and quiescent populations of massive galaxies.

\section{Data Analysis \& SED Fitting}
\label{sec:analysis}
SED fitting is performed using EAZY-py, a Python-based version of EAZY \citep{Brammer2008}, which fits a set of twelve Flexible Stellar Population Synthesis (FSPS; \citealt{Conroy2009}, \citealt{Conroy2010}) templates in non-negative linear combination. We utilize the default set of EAZY-py FSPS templates which are constructed using a \cite{Chabrier2003} initial mass function (IMF), \cite{KriekConroy2013} dust law, solar metallicity, and realistic star-formation histories including bursty and slowly rising models. A full description of the fitting procedure used here and tests of EAZY-py on a set of mock galaxies are given in \cite{Sherman2020} and will briefly be described here. 

For each galaxy in our sample, EAZY-py uses $\chi^2$ minimization to identify the best-fit combination of the twelve built-in templates at the redshift at which $\chi^2$ is minimized, which is the ``best-fit" redshift used throughout this work. We estimate our photometric redshift accuracy using two samples of galaxies with spectroscopic redshifts. The first is a low redshift ($z < 1$) sample of galaxies from the Sloan Digital Sky Survey (SDSS) with spectroscopic redshifts. The second is an intermediate redshift ($1.9 < z < 3.5$) sample of galaxies from the second internal data release of the HETDEX survey \citep{Hill2008}. For each of these samples, we estimate the photometric redshift quality using the normalized median absolute deviation \citep{Brammer2008}:
\begin{eqnarray}
\sigma_{\rm NMAD} &=& 1.48 \times \rm{median}\left( \left| \frac{\Delta z - \rm{median}(\Delta z)}{1 + z_{spec}}  \right| \right).
\end{eqnarray}
We find that for the low redshift sample $\sigma_{\rm NMAD}$ = 0.053 and for the intermediate redshift sample $\sigma_{\rm NMAD}$ = 0.102. We note that the sample used to estimate intermediate redshift photometric redshift recovery is quite small (56 galaxies) and will grow substantially with future data releases from the HETDEX project. Of this small sample from HETDEX, 3 (5.3\%) of the 56 galaxies are catastrophic outliers, where EAZY-py fits them to have $2.0 < z_{\rm phot} < 3.0$ and preliminary spectra from HETDEX suggest that these objects are low-z ($z_{\rm spec} < 0.5$) galaxies. We note that visual inspection was performed for all HETDEX detections to verify the line identification made by the HETDEX pipeline. 

To estimate the stellar masses and star-formation rates for the galaxies in our sample, we implement an improved parameter estimate over that contained in the current EAZY-py procedure. We first use the built-in EAZY-py fitting procedure to find the best fit template combination at the redshift at which $\chi^2$ is minimized. Then, at this redshift we draw 100 SEDs from the best-fit SED's template error distribution \citep{Sherman2020}. Each of these SED draws gives a unique set of galaxy parameters such as stellar mass and SFR, and we construct distributions of these parameters from the 100 draws. When presenting galaxy parameter values throughout this work, we adopt the median of the parameter distribution as the "best-fit" and the 16th and 84th percentiles as the lower and upper error bars, respectively. From this procedure, we find that typical stellar mass and star-formation rate errors in our redshift range of interest ($1.5 < z < 3.0$) are $\pm0.08$ dex and $\pm0.18$ dex, respectively. We note that because the underlying parameters associated with each EAZY-py FSPS template are the intrinsic values for that template, our SFR values for given galaxy SED fits are the extinction corrected intrinsic SFR for that galaxy.  

To validate this SED fitting procedure for redshift and parameter recovery, \cite{Sherman2020} performed extensive tests of EAZY-py by fitting a diverse sample of mock galaxies (V. Acquaviva, private communication). These mock galaxies were assigned realistic error bars using the SHELA footprint sample of galaxies. It was found that EAZY-py does a good job of recovering the underlying galaxy parameters (photometric redshift, stellar mass, and SFR) for this diverse set of mock galaxies, which includes systems with low sSFR. Specifically, for mock galaxies with redshifts $1.5 < z < 3.0$ EAZY-py tends to underestimate the SFR, on average, by 0.46 dex and overestimates stellar mass, on average, by 0.085 dex. These offsets are likely due to systematic differences in the mock galaxy templates constructed from \cite{BruzualCharlot2003} and the FSPS templates used for SED fitting. We note that an under-estimate of the SFR may lead to an over-estimate of the quiescent fraction, and we discuss tests to ensure this is not the case in Section \ref{sec:qf_shela}.

Finally, we perform a test of our SED fitting procedure to estimate both the impact of photometric redshift uncertainty and Eddington bias \citep{Eddington1913} on our results. Eddington bias describes the situation where uncertainties in both photometry and photometric redshifts can cause low-mass galaxies to scatter into high-mass bins. In our test (see also \citealt{Sherman2020}), we generate 100 iterations of our catalog by drawing 100 new photometric redshifts for each galaxy from that galaxy's photometric redshift probability distribution. We then re-fit galaxies at their drawn redshifts using the SED fitting procedure described above and perform the sample selection and analysis presented in Sections \ref{sec_sampleSelection} and \ref{sec:emperical_results} of this work. We find that for $M_\star < 10^{12}$M$_\odot$, the quiescent fractions and stellar mass functions for the 100 iterations are consistent with our empirical results presented in Section \ref{sec:emperical_results}. For $M_\star > 10^{12}$M$_\odot$ we see the effects of Eddington bias, and our results from the 100 catalog iterations differ from our original results by up to a factor of $2-5$. This scatter at the highest masses is expected, and our results throughout this work focus on the mass range $M_\star = 10^{11} - 10^{12}$M$_\odot$. The quiescent fractions from the 100 catalog iterations compared with the results presented in Section \ref{sec:emperical_results} can be found in Appendix \ref{appendix:pertCatalog_test}. This test ultimately shows that photometric redshift uncertainty does not dominate the error on results presented throughout this work. 

We estimate the mass completeness of our sample following the method from \cite{Pozzetti2010} (see also \citealt{Davidzon2013} and \citealt{Sherman2020}) which uses the science sample (see Section \ref{sec_sampleSelection} for sample selection) to estimate completeness. With this method, we scale the mass of the 20\% faintest sample galaxies in small redshift bins such that their $K_s$-band magnitude equals the 95\% completeness limit of our NEWFIRM $K_s$-band survey (22.4 AB mag at $5\sigma$). The 95$^{\rm th}$ percentile of the distribution of scaled galaxy masses in each redshift bin is adopted as the 95\% mass completeness limit for our study for that particular redshift bin. We find that the 95\% mass completeness limits are log($M_{\star}$/$\rm M_{\odot}$) = 10.69, 10.86, and 11.13 in our $1.5 < z < 2.0$, $2.0 < z < 2.5$, and $2.5 < z < 3.0$ bins, respectively. We also performed this procedure using the IRAC 4.5$\mu$m band and found that the 95\% mass completeness values are smaller in each redshift bin. We therefore report the most conservative mass completeness value by using the $K_s$-band. 

Estimating an SFR completeness is not as straightforward as estimating the mass completeness. The SFR values used throughout this work are the extinction-corrected values computed by performing SED fitting with EAZY-py. The SED fitting procedure uses all available data, including the rest-frame UV data (in our case, DECam \textit{u,g,r,i,z}), and an underlying attenuation model to find a best-fit SED and extinction-corrected SFR. Here, we utilize a simplified method to compute a FUV-based SFR and SFR completeness from observed FUV fluxes, which are not corrected for extinction. 

Following \cite{Florez2020} we use the g-band flux as a proxy for FUV-flux. Using a similar procedure to that used for estimating the mass completeness, we begin by identifying the 20\% faintest g-band objects in each of our three redshift bins from redshift $z=1.5$ to 3.0. We then approximate a FUV-based, dust-obscured SFR using the SFR conversion factor from \cite{Hao2011} which assumes a Kroupa IMF \citep{Kroupa2001}, 100 Myr timescale, and a mass range of $0.1 - 100$M$_{\odot}$. We then scale these FUV-based SFR values ($\rm SFR_{FUV}$) to what the SFR would be ($\rm SFR_{lim}$) if that galaxy's g-band magnitude ($\rm m_{gal}$) were the g-band 5$\sigma$ completeness limit ($\rm m_{lim}$ = 24.8 mag AB) found by \cite{Wold2019} using:
\begin{eqnarray}
\log(\rm SFR_{lim}) = \log(\rm SFR_{FUV}) + 0.4(\rm m_{gal} - \rm m_{lim}).
\end{eqnarray}
After this scaling procedure, we find the 95$^{\rm th}$ percentile of the distribution of $\rm SFR_{lim}$, and this value is the 95\% SFR completeness limit for our sample. We find that the 95\% SFR completeness limits are SFR = 3.53, 6.06, and 9.46 M$_{\odot}$ yr$^{-1}$ in our $1.5 < z < 2.0$, $2.0 < z < 2.5$, and $2.5 < z < 3.0$ bins, respectively. We emphasize that the sample used in this work is $K_s$-selected, not g-band selected, so for every galaxy detected in the $K_s$-band our SED fitting procedure fits an extinction-corrected SFR. We do not impose any g-band (or any other rest-UV filter) S/N cut, and therefore, we are able to detect massive galaxies with SFR below the 95\% completeness limit estimated here. 

\subsection{Sample Selection} 
\label{sec_sampleSelection}
Our primary science focus lies in understanding the population of massive quiescent galaxies at cosmic noon ($1.5 < z < 3.0$) in comparison to the total population of galaxies at these redshifts. To achieve this, we place several quality control cuts on our full sample to achieve a high-confidence, representative sample of galaxies. 

We begin by removing all objects identified by SDSS to be stars, spectroscopically confirmed low-z galaxies, and luminous AGN, as well as those identified to be luminous AGN by the Stripe82X survey. This initial cut removes $\sim$4\% of objects from our catalog before SED fitting is performed. Luminous AGN are thought to play an important role in galaxy evolution, however we remove them here as our SED fitting technique inadequately accounts for the contribution from the luminous AGN (e.g., \citealt{Salvato2011}, \citealt{Ananna2017}, \citealt{Florez2020}). A detailed analysis of the properties of galaxies hosting X-ray luminous AGN (with L$_{\rm X} > 10^{44}$ erg s$^{-1}$) in our sample has been performed by \cite{Florez2020} using the CIGALE SED fitting code (\citealt{Noll2009}, \citealt{Ciesla2015}, \citealt{Yang2020}), which performs careful fitting of AGN emission. At $1.5 < z < 3.0$ they find 44 X-ray luminous AGN above their completeness limits and as this sample is quite small compared to our massive galaxy sample, the exclusion of these objects does not significantly impact our results. Previous works (e.g., \citealt{Salvato2011}) have shown that low luminosity AGN (those with $F_{0.5-2 \rm keV}<8\times10^{-15}~ \rm cgs$) are sufficiently well fit with SED templates that do not account for AGN contribution. We do not have a suitable method for identifying or removing low luminosity AGN from our sample, however we believe that they are fit well enough by our SED fitting procedure. 

To ensure adequate SED fits, we require that the best-fit SED has reduced $\chi_{\nu}^2 < 20$ and that the fit is constrained by at least four filters, three of which are NEWFIRM $K_s$, IRAC 3.6 and 4.5$\mu m$. The requirement of four or more filters and $\chi_{\nu}^2 < 20$ removes $\sim2$\% of objects from our sample of objects fit to have redshift $1.5 < z < 3.0$. 

We further require that a galaxy has signal-to-noise S/N $\geq$ 5 in both IRAC 3.6 and 4.5$\mu m$ filters. Our requirement of robust IRAC detections enables the selection of massive galaxies with constraints to their SED fit redward of the Balmer break. \cite{Wold2019} showed that beyond z $\sim$ 1, IRAC data are imperative for reducing photometric redshift error, which is extremely important in this work because we do not place S/N requirements on any other filters. The IRAC S/N requirement further removes $\sim60$\% of objects from our $1.5 < z < 3.0$ sample, leaving 54,001 $K_s$-selected galaxies with robust SED fits and photometric redshifts, of which 28,781 are fit to have $M_\star \ge 10^{11}$M$_\odot$. Of the galaxies with IRAC S/N $<$ 5 that are fit to have $1.5 < z < 3.0$, only  7\% are fit to have $M_{\star} = 10^{11} - 10^{12} M_{\odot}$ (the primary mass range of interest in this work), and 9\% are fit to have $M_\star \ge 10^{12}$M$_\odot$, which is unrealistic given their poor photometric quality. If we were to loosen the strict IRAC S/N $\geq$ 5 criterion and only require IRAC S/N $\geq$ 2, our sample size of $M_\star \ge 10^{11}$M$_\odot$ galaxies would increase to 32,542 galaxies. We choose, however, to use the stricter S/N $\geq$ 5 as this ensures high quality photometric redshift fits. Additionally, we note that employing the looser requirement of IRAC S/N $\geq$ 2 does not change the results presented in this work. 

Due to the low number densities of massive galaxies, a clean sample is important for making conclusions about their nature. At the highest masses (log($M_{\star}$/$\rm M_{\odot}$) $\geq$ 11.5), the most common contaminants are objects fit to have high masses because their light is contaminated by that from a nearby star, typically a bright diffraction spike. To mitigate this type of contaminant, we visually inspect every object fit to have log($M_{\star}$/$\rm M_{\odot}$) $\geq$ 11.5 three times using a custom Zooniverse\footnote{\url{zooniverse.org}} interface. We also adopt visual inspection results for objects in our catalog that were found in the \textit{riz}-selected catalog used by \cite{Sherman2020} who inspected galaxies fit to have $M_\star \ge 10^{11}$M$_\odot$ and found contamination of $\sim$2\% at log($M_{\star}$/$\rm M_{\odot}$) < 11.5. Objects flagged as contaminated by a nearby object are removed from our sample. Following this visual inspection (and adoption of previous inspection results), we are left with a sample of 28,469 $M_\star \ge 10^{11}$M$_\odot$ galaxies with robust masses, star-formation rates, and redshifts. 

\section{Empirical Results}
\label{sec:emperical_results}

\subsection{Empirical Quiescent Fraction As a Function of Mass}
\label{sec:qf_shela}

\begin{figure*}
\begin{center}
\includegraphics[width=5.5in]{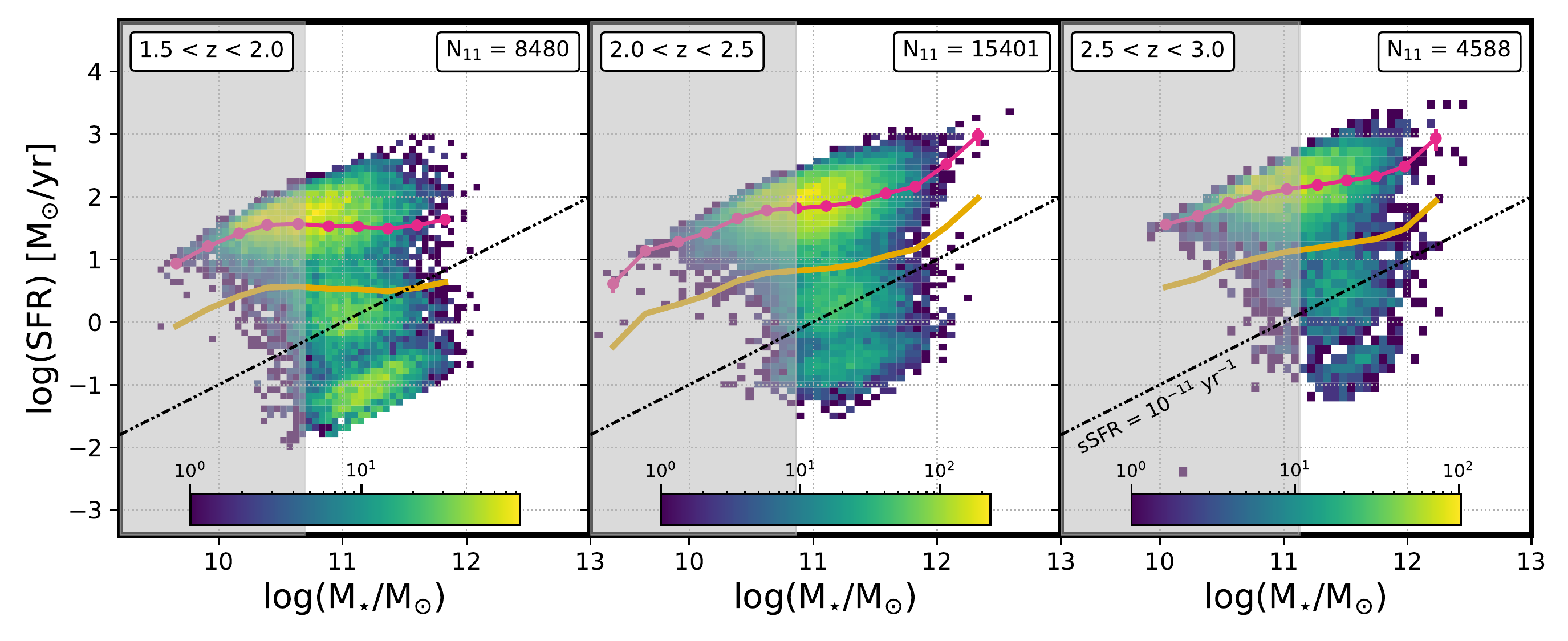} 
\caption{The relationship between star-formation rate (SFR) and stellar mass ($M_\star$) for all galaxies in our sample. The main sequence is represented by dark pink circles, which are the average SFR in a given mass bin (see Section \ref{sec:qf_shela}). Errors on the main sequence are determined using the bootstrap resampling procedure described in Section \ref{sec:qf_shela}. The gold line represents the main sequence - 1 dex which is used in this work to identify quiescent galaxies. The dash-dot line represents the sSFR = 10$^{-11} \rm yr^{-1}$ criterion also used for selecting quiescent galaxies. The main sequence - 1 dex criterion is more effective at selecting green valley galaxies to be quiescent than the sSFR-selection method. Areas of parameter space where our study is not complete in mass are shaded in grey. Inset color bars indicate the number of galaxies in each two dimensional bin. Insets on the upper right of each panel show the number ($N_{11}$) of galaxies in our sample with $M_\star \ge 10^{11}$M$_\odot$.}
\label{ms_shela_plot}
\end{center} 
\end{figure*}  

\begin{figure*}
\begin{center}
\includegraphics[width=5.5in]{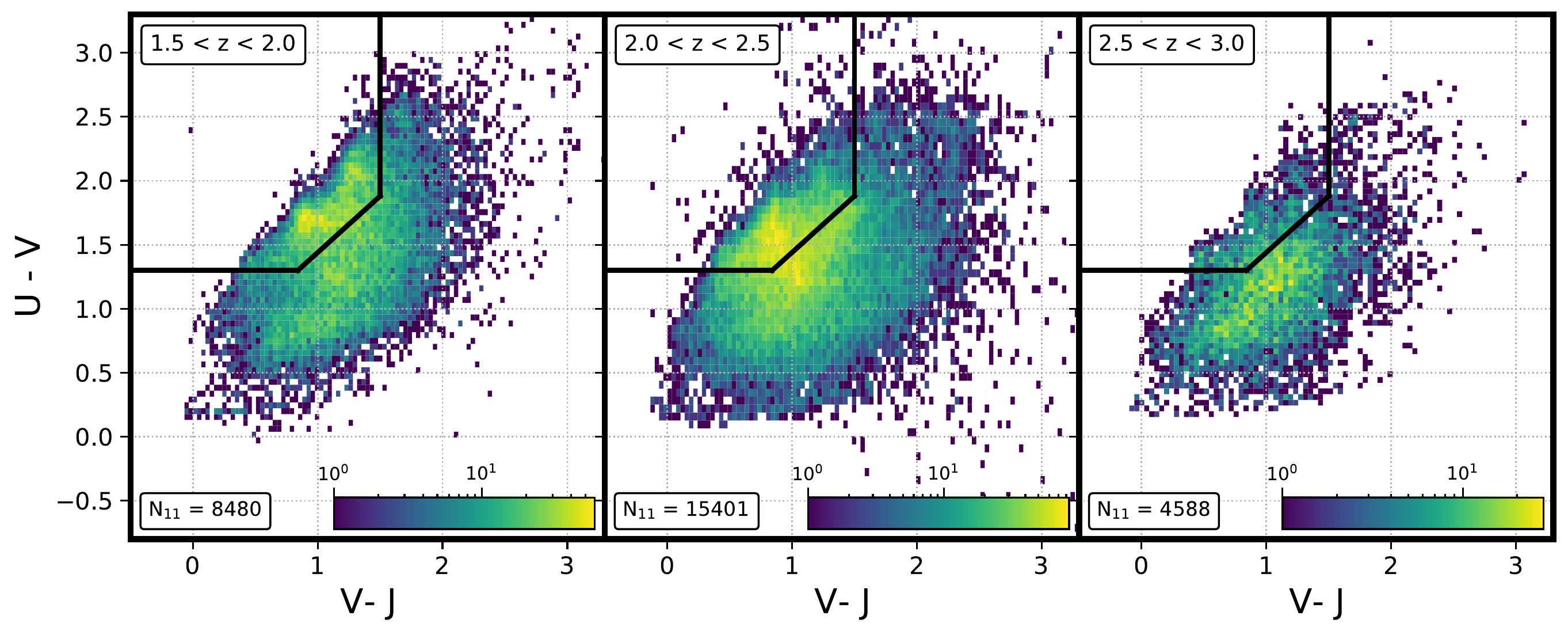} 
\caption{The UVJ color-color diagram for our sample, with the separation of quiescent and star-forming galaxies adopted from \protect\cite{Muzzin2013}. Galaxies in the upper left region of parameter space are quiescent, while those lying to the right of the boundary are dusty star-forming galaxies, and those below the boundary are star-forming galaxies. We see that the population of quiescent galaxies grows from high to low redshift. Inset color bars indicate the number of galaxies in each two dimensional bin. Insets on the lower left of each panel show the number ($N_{11}$) of galaxies in our sample with $M_\star \ge 10^{11}$M$_\odot$.}
\label{uvj_plot}
\end{center} 
\end{figure*}  

When investigating the quiescent population of galaxies in our sample, we utilize three definitions of ``quiescent" to enable fair comparisons with previously published results from observations and different classes of theoretical models. These are:
\begin{enumerate}
\item \textbf{sSFR-Selected:} Galaxies are quiescent when their specific star-formation rate is $\rm sSFR \leq 10^{-11}~\rm yr^{-1}$. \cite{Fontanot2009} implemented this sSFR cut to distinguish between star-forming and quiescent galaxies in a sample of galaxies out to $z\sim4$ using the SFR-$M_\star$ relation to motivate this choice. This definition is straightforward and is simple to compute using the output from SED fitting codes or simulations. This method aims to select galaxies with little recent star-formation, however this threshold ($\rm sSFR \leq 10^{-11}~\rm yr^{-1}$) does not change with redshift. Fixing the sSFR threshold does not account for the higher average star-formation rates of high redshift galaxies compared to local systems with a similar stellar mass. 
\\
\item \textbf{Main Sequence-Selected:} This selection method uses the sample of galaxies to define a main sequence (e.g., \citealt{Tomczak2016}), then identifies any object lying more than 1 dex below that main sequence as quiescent (\citealt{Fang2018}, \citealt{Donnari2019}). This technique is similar to the sSFR-selection method, but allows the threshold for quiescence to vary with redshift and stellar mass. 

The main sequence for our sample of $1.5 < z < 3.0$ galaxies is explored in detail in Sherman et al. (in preparation) and will briefly be described here. We define the main sequence using stellar masses and star-formation rates from our SED fitting analysis using EAZY-py (Fig. \ref{ms_shela_plot}). The value of the main sequence is determined by computing the average SFR in individual mass bins. This method is consistent with the approach from other studies that use mass-complete samples of the total population of galaxies (e.g. \citealt{Whitaker2014}, \citealt{Tomczak2016}). By computing the main sequence in individual mass bins, we leverage our large sample of massive galaxies. We compute the errors on the main sequence by employing a bootstrap procedure. In each bootstrap draw, we select a random sample of galaxies in a given mass bin, where the size of the random sample is equal to the number of galaxies in that mass bin. This is done with replacement, so objects can be selected more than once in a single draw. The bootstrap procedure is repeated 1,000 times, each time computing the average SFR of the random sample to construct a distribution of average SFR values in each mass bin. Lower and upper error bars on the main sequence are the 16$^{\rm th}$ and 84$^{\rm th}$ percentiles of this distribution, respectively. The relationship between SFR and stellar mass, as well as our main sequence is shown in Figure \ref{ms_shela_plot}. The main sequence shown here is in good agreement with that presented by \cite{Tomczak2016}.

When determining the quiescent fraction, we bin our data into the same mass bins used to define the main sequence. In each mass bin, if an object falls 1 dex or more below the main sequence value defined in that bin, it is determined to be quiescent. We find that the value of the main sequence for our $1.5 < z < 3.0$ sample decreases to lower average SFR towards present day (e.g. \citealt{Whitaker2014}, \citealt{Tomczak2016}), and therefore, the threshold for quiescence decreases towards present day. 

In addition to a quiescent threshold that varies with mass and redshift, this method is an improvement over the fixed sSFR threshold as it provides a more meaningful separation of the quiescent and star-forming galaxy populations. In our redshift range of interest ($1.5 < z < 3.0$), the fixed sSFR =  $10^{-11}~\rm yr^{-1}$ threshold runs directly through the so-called green valley \citep{Wyder2007} in the SFR-M$_{\star}$ plane. In contrast, the main sequence-based method effectively separates the main sequence population from the entire green valley population. We note that the quiescent fraction measured from the position relative to the main sequence is dependent on how the main sequence is defined and we take care throughout this work to use only the main sequence definition described here. 
\\
\item \textbf{UVJ-Selected:} With this selection, galaxy rest-frame U, V, and J colors are estimated in EAZY-py based on the shape of the best-fit SED. If the resulting  U - V and V - J colors fall in a particular region of parameter space (Fig. \ref{uvj_plot}), then that galaxy is determined to be quiescent. In this work, we adopt the parameter space used by \cite{Muzzin2013} to select quiescent galaxies in the UVJ plane. 

The bimodality of star-forming and quiescent galaxy populations in the UVJ plane was initially established at high redshift using primarily photometric samples (e.g., \citealt{Labbe2005}, \citealt{Wuyts2007}) and was interpreted using evolutionary tracks from \cite{BruzualCharlot2003} stellar population models. \cite{Williams2009} confirmed the location of quiescent galaxies using a spectroscopically-selected sample of passive galaxies out to $z\sim2$ and implemented dividing lines based on their empirical data. \cite{Muzzin2013} used their sample to update the empirically-based division between these populations. The boundary effectively separates red quiescent galaxies (located in the upper left region of the UVJ diagram) from dusty star-forming galaxies (located in the upper right region of the UVJ diagram) and blue star-forming galaxies (in the lower regions of the UVJ diagram). We note that although rest-frame colors are now standard outputs from SED fitting codes, a limitation of using color to identify quiescent galaxies is that these colors are highly dependent on the stellar population models and dust laws used to fit galaxy photometry. 
\end{enumerate}

\begin{figure*}
\begin{center}
\includegraphics[width=\textwidth]{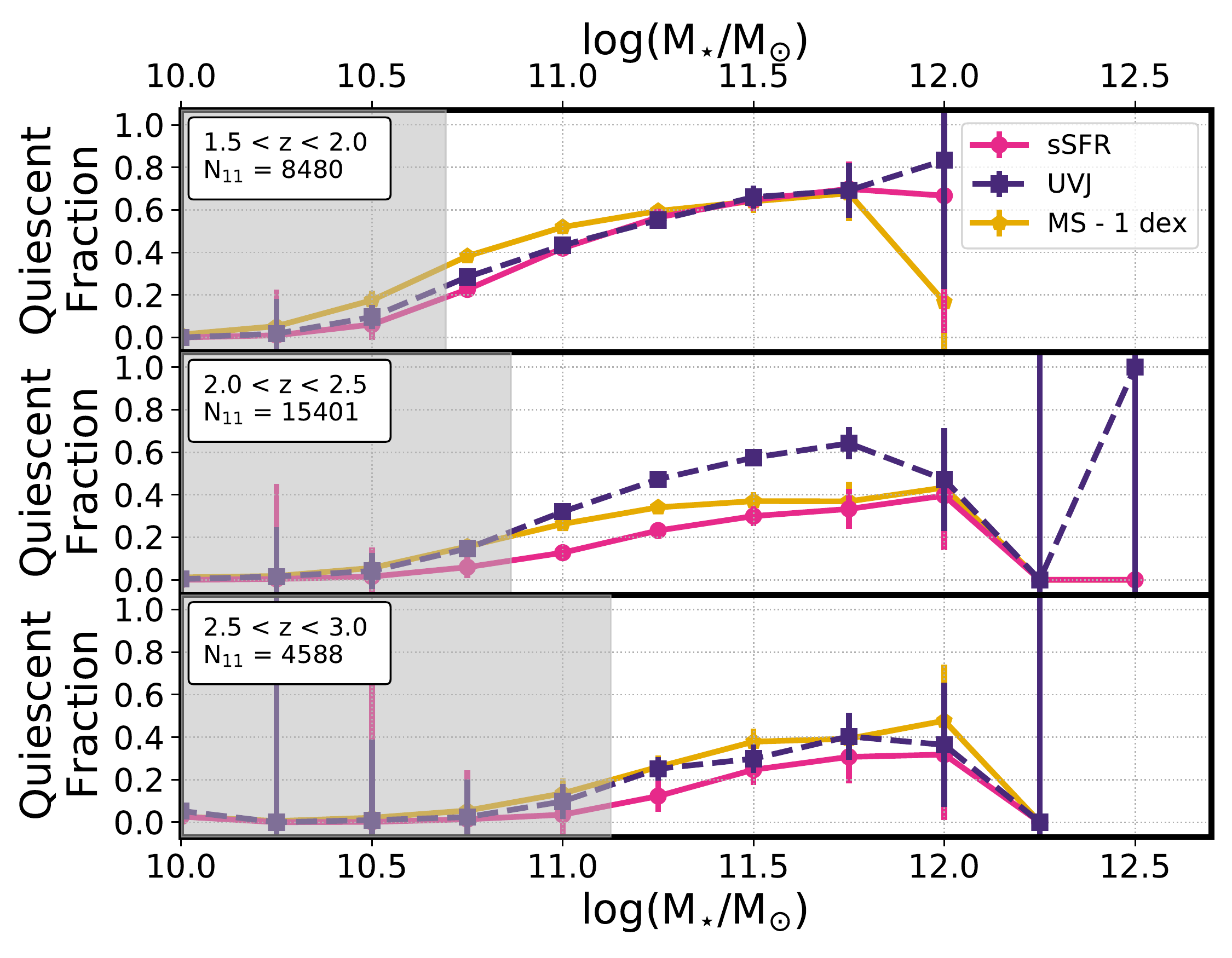} 
\caption{Empirical quiescent fraction for our sample of $1.5 < z < 3.0$ galaxies using the three selection methods described in Section \ref{sec:qf_shela}: sSFR (pink circles), UVJ (purple squares), main sequence - 1 dex (gold pentagons). These three methods of determining the quiescent fraction produce similar trends showing that the quiescent fraction of galaxies increases with mass from $M_{\star} = 10^{11} - 10^{12} M_{\odot}$ in all three redshift bins over the range $1.5 < z < 3.0$ and it also increases as a function of redshift in fixed mass bins. The quiescent fractions measured with these three methods are similar, particularly at $1.5 < z < 2.0$, but differ by up to a factor of 2 at $2.0 < z < 3.0$. It is remarkable that in only 3.3 Gyr (from the Big Bang to $z=2$) the Universe can build and quench more than 25\% of massive ($M_\star = 10^{11}$M$_\odot$) galaxies. Our result is a significant improvement over previous observational studies which used smaller sample sizes (our sample is a factor of 40 larger than that from \protect\cite{Muzzin2013} for log($M_{\star}$/$\rm M_{\odot}$) $\ge$ 11.5 galaxies at these redshifts). The gray shaded region indicates masses below our completeness limit. Insets on the upper left of each panel show the number ($N_{11}$) of galaxies in our sample with $M_\star \ge 10^{11}$M$_\odot$.}
\label{qf_shela_plot}
\end{center} 
\end{figure*} 

We show the quiescent fraction of galaxies in our sample as determined using these three methods in Fig. \ref{qf_shela_plot}. Qualitatively, we find that the quiescent fraction increases from low to high masses and the quiescent fraction increases in fixed mass bins from $z = 3.0$ to $z = 1.5$ regardless of the method used to determine the quiescent fraction. Results are similar in a given redshift bin using all three methods of determining the quiescent fraction with the most significant differences seen in the $2.0 < z < 2.5$ bin where our three methods differ by up to a factor of 2. 

In Section \ref{sec:analysis} we noted that \cite{Sherman2020} found that EAZY-py, the SED fitting code used in this work, may systematically under-estimate the SFR by 0.46 dex, on average, which would potentially lead to an over-estimate of the quiescent fraction. We do not believe this is the case as the sSFR-based quiescent fraction is the only one that would be affected by a systematic under-estimate of the SFR and it is in good agreement with the UVJ and main sequence-based quiescent fractions, which are not impacted by a systematic under-estimate of the SFR. The UVJ color-color method would not be affected as it uses the shape of the underlying SED to extract galaxy rest frame colors. The main sequence-based method separates star-forming and quiescent galaxies based on their relative distance from the main sequence (which is defined using the sample of galaxies) and would therefore not be affected. Of our three methods, the only one that would be impacted by a systematic under-estimate of the SFR is the sSFR-based method, which employs a fixed cutoff to separate star-forming and quiescent galaxies. We performed a test where we increase the SFR for all galaxies in our sample by 0.46 dex and recompute our sSFR-based quiescent fraction. With the systematic increase in the SFR, our sSFR-based quiescent fraction would decrease by less than a factor of 2, and we would still find that the quiescent fraction increases as a function of stellar mass. The results presented throughout this work would not differ significantly if we were, in fact, systematically under-estimating the SFR by 0.46 dex.

As an example of our quiescent fraction results, using the main-sequence based quiescent fraction we find that at early epochs ($2.5 < z < 3.0$) the quiescent fraction increases from $13.5\%\pm7.1\%$ at log($M_{\star}$/$\rm M_{\odot}$) = 11 to $39.6\%\pm11.2\%$ at log($M_{\star}$/$\rm M_{\odot}$) = 11.75. At intermediate epochs ($2.0 < z < 2.5$) the quiescent fraction increases from $26.3\%\pm2.5\%$ at log($M_{\star}$/$\rm M_{\odot}$) = 11 to $36.7\%\pm9.1\%$ at log($M_{\star}$/$\rm M_{\odot}$) = 11.75. It is remarkable that by $z=2$ (only 3.3 Gyr after the Big Bang) the universe has managed to quench more than 25\% of massive ($M_\star = 10^{11}$M$_\odot$) galaxies. At later epochs ($1.5 < z < 2.0$) the main sequence-based quiescent fraction is $51.9\%\pm2.5\%$ at log($M_{\star}$/$\rm M_{\odot}$) = 11 and increases to $66.4\%\pm13.1\%$ at log($M_{\star}$/$\rm M_{\odot}$) = 11.75. The two other methods by which we measure the quiescent fraction (sSFR- and UVJ-based) give similar results. 

\subsection{Estimating Contamination From DSFGs}
\label{sec:dsfg}
Dusty star-forming galaxies (DSFGs) are a population of galaxies that require careful consideration because their star-formation is known to be highly obscured by dust (e.g., \citealt{Papovich2006}, \citealt{Casey2014} and references therein, \citealt{Escalante2020}), particularly at $z\sim2$. Here, we explore the degree to which our sample of quiescent galaxies might be contaminated with DSFGs by using a subset of our sample that has available far-IR data.

A  robust way of identifying DSFGs is to use long-wavelength data, in our case \textit{Herschel}-SPIRE (HerS, \citealt{Viero2014}) far-IR/submillimeter taken at 250, 350, and 500$\mu$m. These data are capable of breaking the degeneracy because a HerS detection indicates that photons produced during star-formation are absorbed by dust and reemitted in the far-IR. Although the HerS data has been taken across the majority of our survey footprint, the resolution is poor (the 250$\mu$m band has 18$\arcsec$ resolution) compared to our lower-wavelength data. Because of this, careful consideration must be taken for blended objects as it is difficult to disentangle the contribution from several nearby $K_s$-selected objects. To eliminate this issue, we only utilize isolated HerS objects in our contamination estimate. 

To select a sample, we first identify all massive ($M_\star \ge 10^{11}$M$_\odot$) $K_s$-selected objects fit to have $1.5 < z < 3.0$ that are position matched to the HerS catalog and have positive flux values in all three HerS bands. There are 222 such objects in our catalog. Next, we match the $K_s$-selected objects that have HerS matches to the full $K_s$-selected catalog in the SHELA footprint. If a $K_s$-selected object with HerS detections has another $K_s$-selected object within 9$\arcsec$, then that object is discarded. After this procedure, we are left with 29 massive $1.5 < z < 3.0$ $K_s$-selected galaxies with isolated HerS detections. We emphasize that the HerS data is \textit{not} used in our SED fitting procedure because we are simply using the HerS data to classify these 29 galaxies as DSFGs. We do not aim to correct SED fits to get the obscured SFR.

Because a HerS detection implies that a galaxy is a DSFG, if any of our three methods identify these galaxies to be quiescent, then they should be considered a contaminant to the quiescent sample. For the sSFR-based method, 6 (26\%) of the isolated HerS sample are fit by EAZY-py to be quiescent. Using the UVJ-based method we find that 1 (3.5\%) galaxy in our sample of HerS-identified DSFGs is placed into the quiescent population, and for the main sequence-based method 8 (38\%) of our DSFGs are labeled as quiescent. With such a small sample of DSFGs, it is difficult to provide definitive estimates of the contamination due to DSFGs, however it is encouraging that the UVJ-based method of determining the quiescent fraction is largely uncontaminated and this method provides similar empirical quiescent fraction results to our other two methods. 

\subsection{Comparing the Empirical Quiescent Fraction to Previous Observations}
\label{sec:shela_qf_v_obs}

\begin{figure*}
\begin{center}
\includegraphics[width=5in]{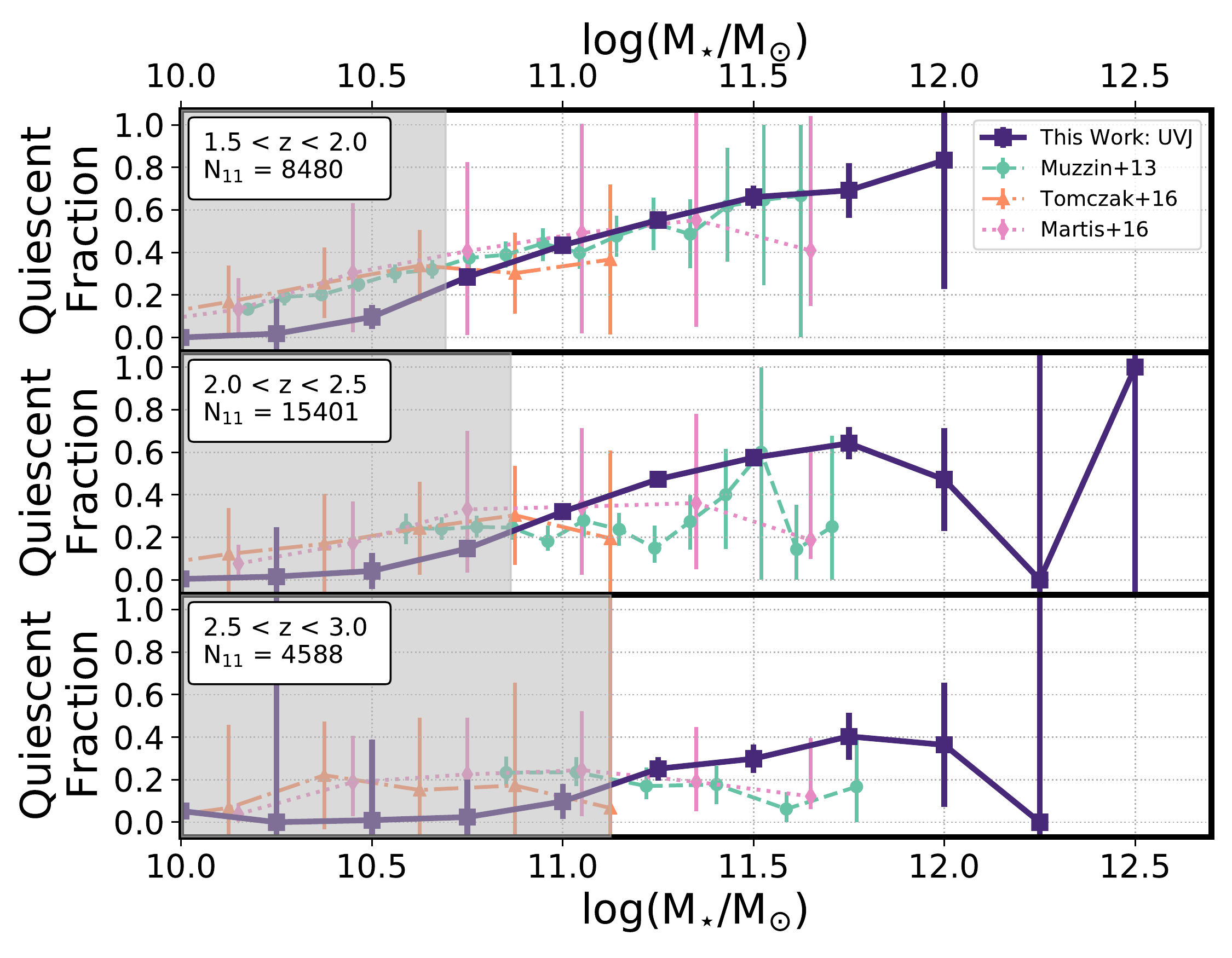} 
\caption{Empirical UVJ-selected quiescent fraction (purple squares) compared with previous observational results from \protect\cite{Muzzin2013} (green circles), \protect\cite{Tomczak2016} (orange triangles), and \protect\cite{Martis2016} (pink diamonds). Our work extends to higher masses and our larger sample size allows us to achieve smaller errors than previous works. Our sample of galaxies with log($M_{\star}$/$\rm M_{\odot}$) $\ge$ 11.5 is a factor of $\sim$40 larger than samples from previous works. The quiescent fraction measured from our sample is a factor of $\sim2-3$ larger than that from previous studies at $2.0 < z < 3.0$ and we find good agreement at $1.5 < z <2.0$, however the errors from previous studies are large. The gray shaded region indicates masses below our completeness limit. Insets on the upper left of each panel show the number ($N_{11}$) of galaxies in our sample with $M_\star \ge 10^{11}$M$_\odot$.}
\label{qf_obs_uvj_fig}
\end{center} 
\end{figure*}  

We compare our empirical quiescent fraction as a function of mass with quiescent fraction measures from several previous studies (Fig. \ref{qf_obs_uvj_fig}; \citealt{Muzzin2013}, \citealt{Martis2016}, \citealt{Tomczak2016}) that use the UVJ-selection method. The previous works all implement similar boundaries between quiescent and star-forming populations to those adopted in our study from \cite{Muzzin2013}. 

As outlined in Section \ref{sec:qf_shela}, our study finds that the quiescent fraction of massive galaxies ($M_\star > 10^{11}$M$_\odot$) measured in three ways increases as a function of mass from $M_{\star} = 10^{11} - 10^{12} M_{\odot}$ in all three redshift bins over the range $1.5 < z < 3.0$ and it also increases as a function of redshift, towards present day, in fixed mass bins. Using the UVJ-based selection method, we find that at early epochs ($2.5 < z < 3.0$) the quiescent fraction is $9.7\%\pm8.2\%$ at log($M_{\star}$/$\rm M_{\odot}$) = 11 and increases with mass to $40.4\%\pm11.1\%$ at log($M_{\star}$/$\rm M_{\odot}$) = 11.75. At later epochs ($1.5 < z < 2.0$), we find that the UVJ-based method gives a quiescent fraction of $43.4\%\pm2.7\%$ at log($M_{\star}$/$\rm M_{\odot}$) = 11, which increases with mass to $69.1\%\pm12.9\%$ at log($M_{\star}$/$\rm M_{\odot}$) = 11.75. 

Our work significantly extends that from previous studies to higher stellar masses at $1.5 < z < 3.0$ due to the larger volume probed by our 17.5 deg$^2$ study compared with \cite{Muzzin2013} (1.62 deg$^2$), \cite{Martis2016} (1.62 deg$^2$), and \cite{Tomczak2016} (400 arcmin$^2$). Our sample of galaxies with log($M_{\star}$/$\rm M_{\odot}$) $\ge$ 11.5 is a factor of $\sim$40 larger than samples from \cite{Muzzin2013}, \cite{Martis2016}, and \cite{Tomczak2016}. This larger sample size allows for significantly smaller error bars than previous studies and our larger volume renders errors from cosmic variance negligible in our work (cosmic variance for redshift $1.5 < z < 3.0$ $M_\star \ge 10^{11}$M$_\odot$ galaxies is $\sim10-30\%$ for the samples from \cite{Muzzin2013} and \cite{Martis2016}, and $\sim50-70\%$ for the sample from \cite{Tomczak2016}). 

Direct comparisons with previous results are challenging at the high mass end due to the small sample sizes and large error bars from previous works. We do, however, note some distinct trends. At redshifts $2.0 < z < 3.0$ our empirical quiescent fraction is a factor of $\sim2-3$ higher at  $M_\star > 10^{11}$M$_\odot$ than the quiescent fraction found by earlier studies. In our lowest redshift bin ($1.5 < z < 2.0$) there is general agreement in the quiescent fraction at the high mass end, although the error bars from previous studies remain quite large. 

Differences in the measured quiescent fraction among observational studies are not only attributable to sample size, but also the stellar population models and dust laws used in SED fitting. Rest-frame colors are typically estimated by passing the best-fit SED from the SED fitting procedure through the U, V, and J filter transmission curves. Therefore, systematic differences in the stellar population models used for SED fitting can lead to systematic differences in the measured rest-frame UVJ colors and, subsequently, the quiescent fraction measured from these colors. 

\subsection{Empirical Quiescent Fraction As a Function of Redshift}
\label{sec:qf_shela_vz}
\begin{figure}
\begin{center}
\includegraphics[width=3.3in]{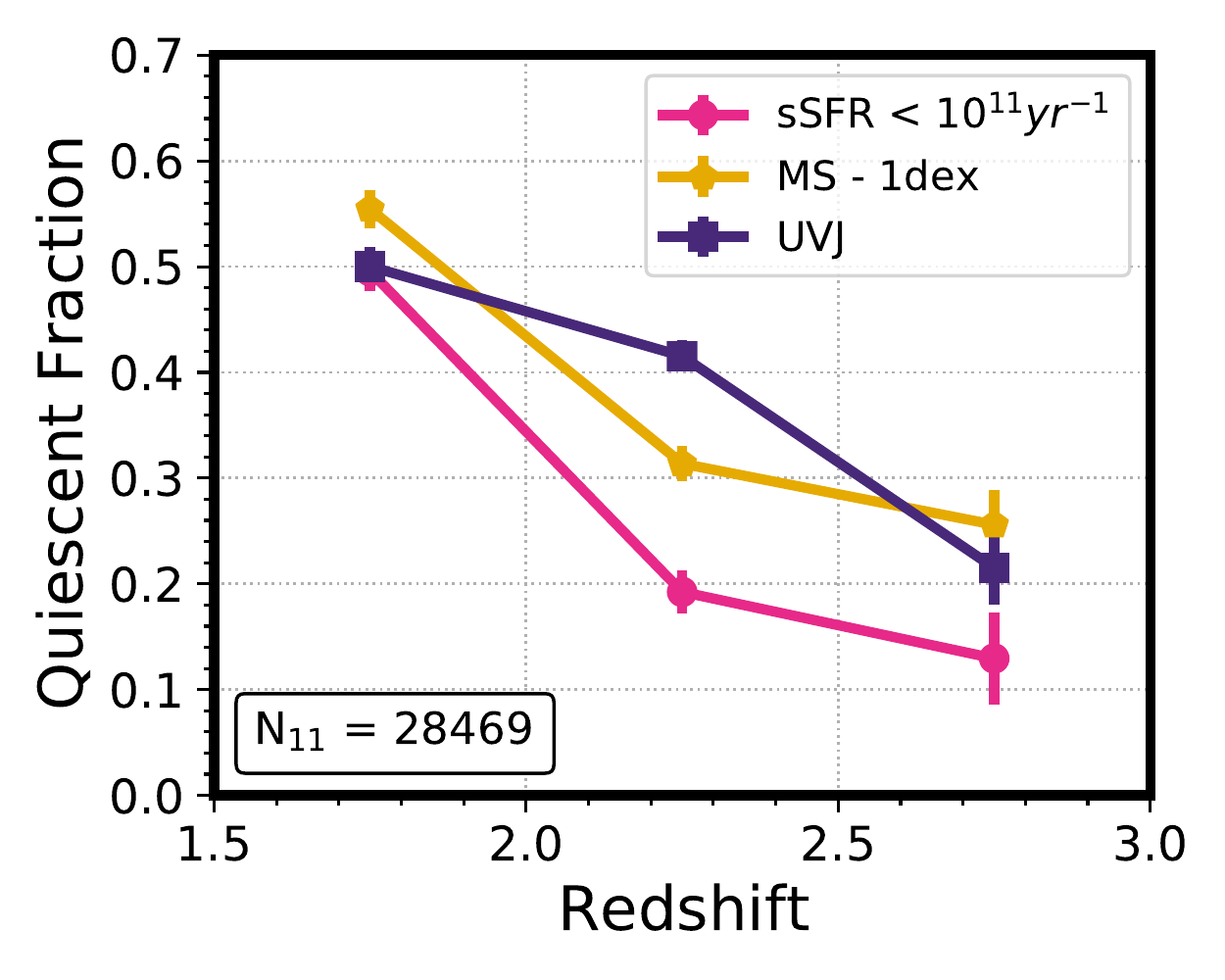} 
\caption{Quiescent fraction for galaxies in our sample with $M_\star \ge 10^{11}$M$_\odot$ shown for our three redshift bins. We show our results using three different methods of computing the quiescent fraction and find that all three methods give a quiescent fraction that increases from high to low redshift. In our highest redshift bin ($2.5<z<3.0$), we find quiescent fractions that span from 13\% (sSFR-based selection) to 26\% (main sequence-based selection). In our lowest redshift bin ($1.5<z<2.0$) our empirical quiescent fractions span from 50\% (sSFR and UVJ-based selection) to 55\% (main sequence-based selection). The inset on the lower left shows the number ($N_{11}$) of galaxies in our sample with $M_\star \ge 10^{11}$M$_\odot$ across all three redshift bins spanning $1.5 < z < 3.0$.}
\label{qf_z_func}
\end{center} 
\end{figure}

\begin{figure*}
\begin{center}
\includegraphics[width=\textwidth]{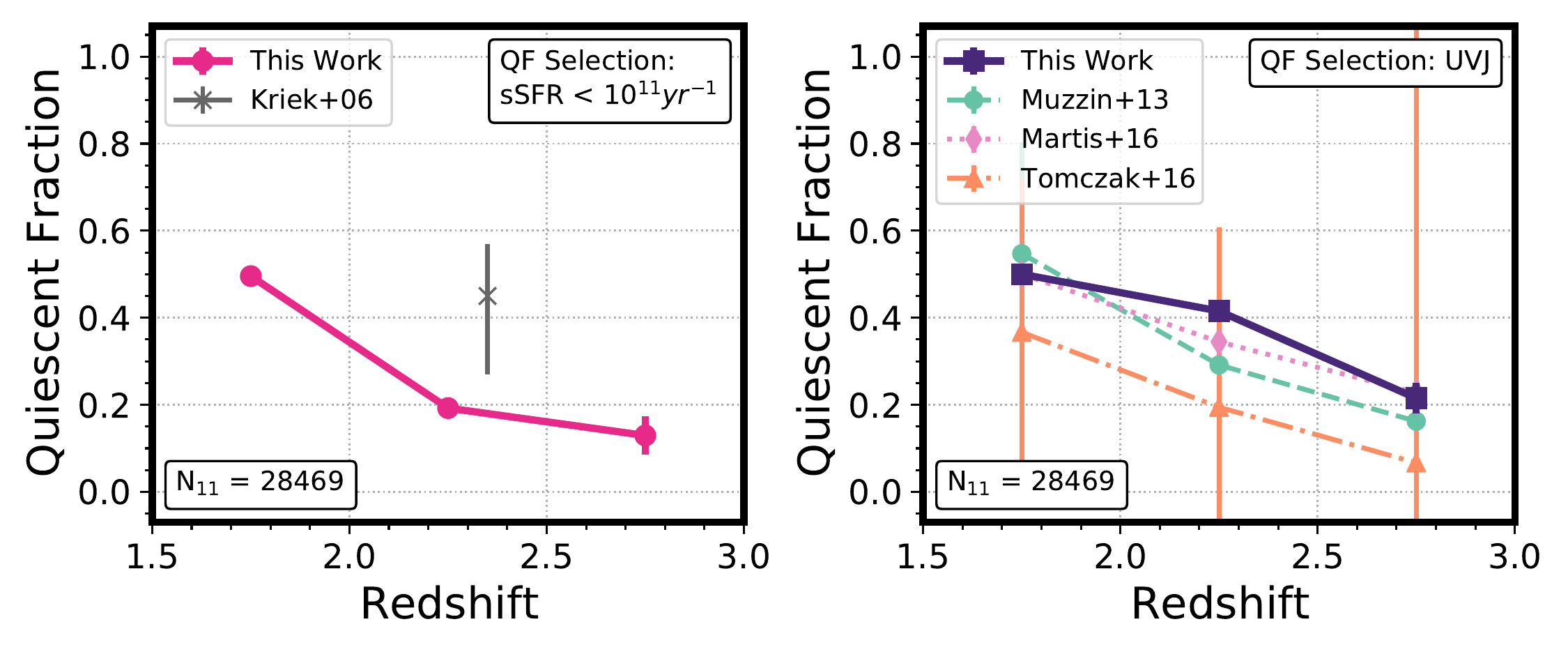} 
\caption{Our empirical quiescent fraction for all massive ($M_\star \ge 10^{11}$M$_\odot$) galaxies selected using the sSFR (left) and UVJ (right) methods as a function of redshift compared with previous observations. For the sSFR-based method, \protect\cite{Kriek2006} finds the quiescent fraction at $z\sim2.5$ to be higher than our empirical result by a factor of $\sim$2. Our UVJ-based result is in good agreement with those from \protect\cite{Muzzin2013} and \protect\cite{Martis2016}, however our empirical quiescent fraction is larger than that from \protect\cite{Tomczak2016} by a factor of 2. Insets on the lower left of each panel show the number ($N_{11}$) of galaxies in our sample with $M_\star \ge 10^{11}$M$_\odot$ across all three redshift bins spanning $1.5 < z < 3.0$. Our sample is more than an order of magnitude larger than samples from previous studies, which allows for smaller Poisson errors.}
\label{qf_z_func_vLit}
\end{center} 
\end{figure*}

In the previous sections, we focused on our results as a function of mass as that highlights the major strengths of our study (large, uniformly selected sample with a statistically significant number of massive galaxies). Here, we show the quiescent fractions in our sample using the three methods presented throughout this work, as a function of redshift for galaxies with $M_\star \ge 10^{11}$M$_\odot$ (Fig. \ref{qf_z_func}). Additionally, we present comparisons with previous observational results (Fig. \ref{qf_z_func_vLit}). 

When we explore the quenched fraction (computed for all galaxies in our sample with $M_\star \ge 10^{11}$M$_\odot$) as a function of redshift, we find that the sSFR-selection method produces the lowest quiescent fraction for all massive galaxies in every redshift bin. The UVJ-selection method produces the highest quiescent fraction for all massive galaxies in the $2.0 < z < 2.5$ bin, while the main sequence-based technique gives the highest quiescent fraction for all massive galaxies in the $1.5 < z < 2.0$ and $2.5 < z < 3.0$ bins. Regardless of the method used to compute the quiescent fraction for all massive galaxies, we find that the quiescent fraction increases from high redshift (quiescent fraction spanning $\sim$ 13 - 26\% across our three methods of estimating the quiescent fraction) to low redshift (quiescent fraction spanning $\sim$ 50 - 55\% across our three methods of estimating the quiescent fraction). 

A key advantage of our study is that our statistically significant sample selected over a large area gives us small errors which are dominated by Poisson error (error from cosmic variance is negligible). Our large sample size allows us to robustly establish that the quenched fraction for all galaxies with stellar masses $M_\star \ge 10^{11}$M$_\odot$ increases from high to low redshift. As an example, we find that with the main-sequence based method the quiescent fraction at the massive end increases from $25.6\%\pm3.3\%$ at $2.5 < z < 3.0$ to $55.4\%\pm1.8\%$ at $1.5 < z < 2.0$. Earlier studies (\citealt{Muzzin2013}, \citealt{Martis2016}, \citealt{Tomczak2016}) with much smaller samples found a similar trend with redshift (see Fig. \ref{qf_z_func_vLit}), albeit with much larger error bars. 

\subsection{Empirical Stellar Mass Function for Star-Forming, Quiescent, and All Galaxies}
\label{sec:shela_smf}
Like the quiescent fraction of massive galaxies, the galaxy stellar mass function is an important tool in understanding the way in which massive galaxies evolve. While the quiescent fraction provides insights about the processes that quench star-formation in the massive galaxy population, the stellar mass function gives information about the buildup of the entire massive galaxy population per unit volume. The slope and normalization of the galaxy stellar mass function at the high mass end place important constraints on theoretical models.

\begin{figure*}
\begin{center}
\includegraphics[width=5.5in]{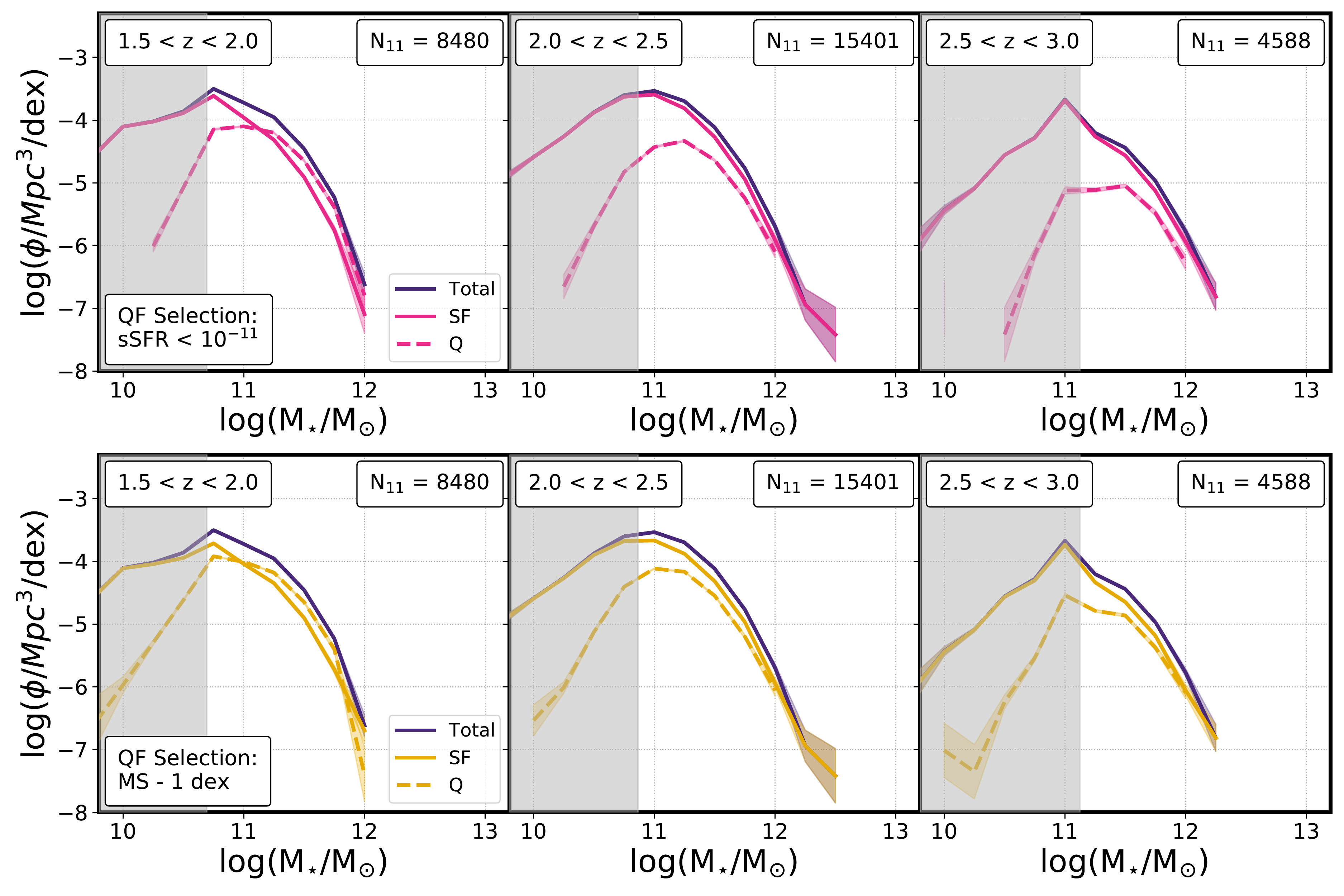} 
\caption{The empirical galaxy stellar mass function for our sample of massive galaxies. In both the top and bottom rows, the purple line represents the galaxy stellar mass function for all galaxies in our sample. In the top row, the solid and dashed pink lines are the star-forming and quiescent galaxy stellar mass functions, respectively, where the quiescent galaxies were selected using the sSFR-based method. Similarly, in the bottom row the gold solid and dashed lines are the star-forming and quiescent galaxy stellar mass functions, respectively, with quiescent galaxies selected by the main sequence-based method. Poisson errors are indicated by the colored regions and are often smaller than the lines. The total, star-forming, and quiescent galaxy stellar mass functions are related through the quiescent fraction, as described in Equation \ref{eqn:smf}. In each of our three redshift bins spanning $1.5 < z < 3.0$, we find the stellar mass function to be steeply declining at the high mass end. The gray shaded region indicates masses below our completeness limit. Insets on the upper right of each panel show the number ($N_{11}$) of galaxies in our sample with $M_\star \ge 10^{11}$M$_\odot$.}
\label{smf_shela}
\end{center} 
\end{figure*} 

\begin{figure*}
\begin{center}
\includegraphics[width=5.5in]{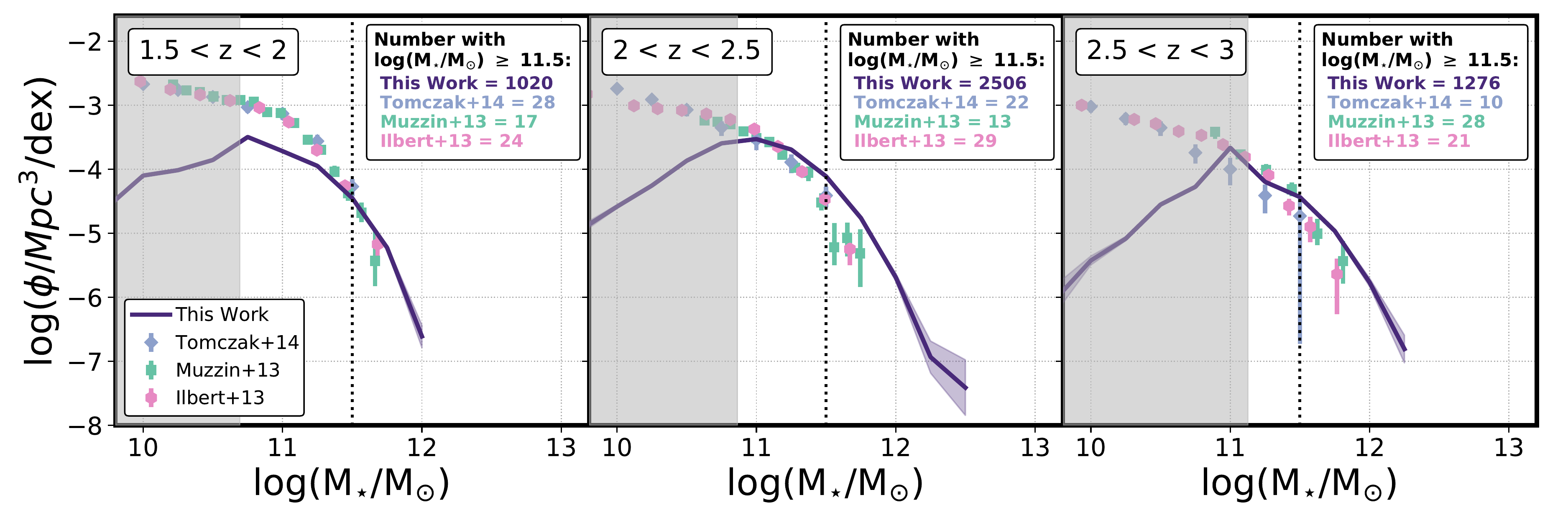} 
\caption{The empirical galaxy stellar mass function for our total sample of massive galaxies (purple line) compared with the total galaxy stellar mass functions from previous works (\protect\citealt{Ilbert2013}, \protect\citealt{Muzzin2013}, and \protect\citealt{Tomczak2014}). The gray shaded region indicates masses below our completeness limit. We find fair agreement, within a factor of $\sim2-3$, with previous results in the two redshift bins spanning $2.0 < z < 3.0$ (center and right panels). In the lowest redshift bin, our result is a factor of $\sim3$ lower at log($M_{\star}$/$\rm M_{\odot}$) $< 11.2$ than previous studies (see discussion in Section \ref{sec:shela_smf}). Poisson errors for our empirical result are indicated by the light purple regions and are often smaller than the lines. Insets on the upper right of each panel show the number of galaxies in our sample and those we compare with that have log($M_{\star}$/$\rm M_{\odot}$) $> 11.5$. The dotted vertical line marks log($M_{\star}$/$\rm M_{\odot}$) $= 11.5$. Our sample is more than a factor of 40 larger than samples from previous studies. }
\label{smf_v_obs}
\end{center} 
\end{figure*} 

\cite{Sherman2020} explored the galaxy stellar mass function for a sample of star-forming galaxies and compared their results with those from previous observational studies and theoretical models. As was discussed earlier, the sample from \cite{Sherman2020} was taken in the same footprint as the sample used in this work. However, since the previous work was \textit{riz}-selected the sample was biased towards star-forming galaxies. With the $K_s$-band selection used in this work, we are now able to explore the high mass end of the stellar mass function for all galaxies, those that are star-forming, and those that are quiescent.

Our stellar mass function is computed using the $1/V_{\rm max}$ method \citep{Schmidt1968} following the procedure of \cite{Weigel2016} (see also \citealt{Sherman2020}). We begin by using our procedure for estimating the 95\% mass completeness (Section \ref{sec:analysis}) to find the redshift at which our study is 95\% complete in a given mass bin. We then split our sample into three redshift bins ($1.5 < z < 3.0$, $\Delta z = 0.5$) and further bin by mass. For each mass bin, the accessible volume is computed using the maximum redshift ($z_{\rm max}$) at which our study is 95\% complete in that mass bin. If $z_{\rm max}$ is greater than the maximum redshift of a given redshift bin, then the comoving volume is set to be the volume of the redshift bin of interest. In the case that $z_{\rm max}$ lies within the redshift bin of interest, the comoving volume is measured to be the volume between the minimum redshift of that bin and $z_{\rm max}$. In the event that $z_{\rm max}$ is less than the minimum redshift of a given bin (this only occurs for mass bins well below our 95\% mass completeness limit), then the comoving volume is set to be the minimum of the comoving volume of the redshift bin of interest or the comoving volume from $z = 0$ to $z_{\rm max}$. 

The procedure described here is used to compute the stellar mass function for our total sample of galaxies as well as star-forming and quiescent samples. Throughout this work we have used three methods to separate star-forming and quiescent galaxies: sSFR, distance from the main sequence, and UVJ. Our interest lies in later comparing our results with those from theoretical models (see Section \ref{sec:theory_smf}), for which we only separate star-forming and quiescent galaxies using the sSFR-based and main sequence-based methods. Therefore, we employ only the sSFR and main sequence-based methods for separating star-forming and quiescent galaxies. These two methods of separating star-forming and quiescent galaxies give similar star-forming and quiescent galaxy stellar mass functions, which is not surprising since both of these methods give similar quiescent fractions as a function of mass. 

In Figure \ref{smf_shela} we show the total stellar mass function and star-forming and quiescent populations split using the sSFR-based method and the main-sequence based method (top and bottom rows of Fig. \ref{smf_shela}, respectively). The total galaxy stellar mass function ($\Phi_{\rm tot}$) is related to the star-forming galaxy stellar mass function ($\Phi_{\rm SF}$) and quiescent galaxy stellar mass function ($\Phi_{\rm Q}$) through the quiescent fraction ($f_{\rm Q}$; see Section \ref{sec:qf_shela} and Fig. \ref{qf_shela_plot}) as follows:
\begin{eqnarray}
\label{eqn:smf}
\Phi_{\rm tot} = \Phi_{\rm Q} + \Phi_{\rm SF}  = f_{\rm Q} \times \Phi_{\rm tot} + \Phi_{\rm SF}
\end{eqnarray}
We find for the total, star-forming, and quiescent stellar mass functions that the stellar mass function is steeply declining at the high mass ($M_\star \ge 10^{11}$M$_\odot$) end. 

We also compare our total galaxy stellar mass functions with results from \cite{Ilbert2013}, \cite{Muzzin2013}, and \cite{Tomczak2014} (Fig. \ref{smf_v_obs}). Above our stellar mass completeness limit in the two higher redshift bins ($2.0 < z < 2.5$ and $2.5 < z < 3.0$), our stellar mass function is in fair agreement, within a factor of $\sim2-3$ with previous studies, which have large error bars at high masses due to small number statistics.  However, in the lowest redshift bin ($1.5 < z < 2.0$), our stellar mass function is lower by a factor of $\sim3$ than previous results at masses log($M_{\star}$/$\rm M_{\odot}$) $< 11.2$. Our tests suggest that in this redshift bin, our filter coverage is not as sensitive to SED features (such as the Balmer break and UV slope), which may lead to the deficit of galaxies in this bin. We note, however, that cosmic variance significantly impacts the studies we compare with by $10-30\%$ for \cite{Ilbert2013} and \cite{Muzzin2013} and $50-70\%$ for \cite{Tomczak2014}, while it is negligible for our study due to the large area covered by our data. The significant impact of cosmic variance on previous studies may drive the difference seen between our result and results from previous works, and we treat the stellar mass function in this lowest redshift bin with caution throughout this paper. We will compare our stellar mass function results to different classes of theoretical models in Section \ref{sec:theory_smf} to evaluate how well the models predict both the quenched fraction and the overall population of massive galaxies.

\section{Discussion}
\label{sec:discussion}
In previous sections, we showed that the quiescent fraction of massive galaxies increases as a function of mass (Section \ref{sec:qf_shela}) and redshift (Section \ref{sec:qf_shela_vz}). In Section \ref{sec:quench_mech}, we discuss which physical processes may contribute to this finding across different epochs and environments. We then compare our empirical quiescent fraction and stellar mass function results with several classes of theoretical models in Sections \ref{sec:qf_mass_vTheory}, \ref{sec:qf_z_vTheory}, and \ref{sec:theory_smf}. These comparisons provide key benchmarks for these models and can be used to implement future improvements. 

\subsection{Quenching Mechanisms Across Different Epochs and Environments}
\label{sec:quench_mech}
Our study allows for one of the most robust investigations to date of the buildup of the quiescent galaxy population as a function of mass at the highest masses ($M_\star \ge 10^{11}$M$_\odot$) at cosmic noon ($1.5 < z < 3.0$). This is achieved due to our large sample size (28,493 galaxies with $M_\star \ge 10^{11}$M$_\odot$) selected over a 17.5 deg$^2$ area, which gives small errors dominated by Poisson statistics and renders errors from cosmic variance negligible. We showed that the quiescent fraction computed in three different ways rises with stellar mass from $M_\star = 10^{11}$ to $10^{12}$M$_\odot$ in three redshift bins spanning $1.5 < z < 3.0$ (Fig. \ref{qf_shela_plot}). Additionally, we show that the quiescent fraction of massive galaxies increases towards present day (Fig. \ref{qf_shela_plot} and Fig. \ref{qf_z_func}). 

To interpret these results, we first consider the population of massive galaxies at high redshifts ($z\sim3$). At early epochs ($z=3$ is only 2.2 Gyr after the Big Bang) massive galaxies are believed to have stemmed from density fluctuations that grow hierarchically through gravitational instability (e.g., \citealt{Springel2005}). These overdensities may evolve into proto-clusters (e.g., \citealt{Lotz2013}, \citealt{Overzier2016}, \citealt{Chiang2017}) and proto-groups \citep{Diener2013}, which are the likely progenitors of modern day clusters and groups. Galaxy interactions and mergers are common in these overdense environments, which are conducive to rapid growth and mass buildup.

\cite{Chiang2017} used results from the Millennium Simulation (\citealt{Springel2005}, \citealt{Guo2013}, \citealt{Henriques2015}) to show that these proto-clusters, while diffuse and rare in number density compared to their modern day descendants, are responsible for $\sim$30\% of the cosmic star-formation rate density at $z\sim3$. They also report that the cores of these proto-clusters are home to only $\sim$30\% of the mass and star-formation within the proto-cluster, indicating that massive galaxies residing in proto-clusters are not necessarily the central galaxy at early times. These results suggest that massive galaxies in proto-clusters at $z\sim3$ (and even to $z\sim1$) move with respect to the proto-cluster core and are subject to environmental effects, such as tidal interactions and major or minor mergers.

Galaxy-galaxy interactions are frequent in proto-cluster environments. Unlike in present day clusters, proto-cluster environments (which are similar environments to present-day groups) have high galaxy number densities and low velocity dispersions; conditions which often lead to mergers. Mergers are favored when the galaxy velocity dispersion within the group is smaller than the average stellar velocity within the interacting galaxies \citep{BinneyTremaine1987}. Major and minor mergers generate large gas inflows to the central regions of the galaxy which increase central gas densities and enhance star-formation rates (e.g., \citealt{Hernquist1995}, \citealt{Mihos1996}, \citealt{DiMatteo2007}, \citealt{Jogee2009}, \citealt{Robaina2010}). These high star-formation rates cause massive galaxies to use their gas supply faster and quench at early epochs. As major mergers trigger central starbursts and AGN activity (e.g., \citealt{Springel2005}, \citealt{Jogee2006} and references therein, \citealt{DiMatteo2008}, \citealt{Capelo2015}, \citealt{Park2017}), stellar and AGN feedback can heat, expel, and redistribute gas, which suppresses future star-formation. Therefore, at early epochs, frequent mergers coupled with stellar and AGN feedback can act as a powerful quenching mechanism, first accelerating, then suppressing star-formation. We note that without spectroscopic redshifts or high resolution imaging, we are unable to accurately determine a merger rate for our sample. 

As galaxies evolve, they accrete gas from the ionized intergalactic medium at all epochs. The gas is accreted through both the hot mode where gas is shock heated to the virial temperature of the halo, as well as the cold mode where gas is fed via cold, dense intergalactic filaments that penetrate the halo without shock heating (e.g., \citealt{Birnboim2003}, \citealt{Katz2003}, \citealt{Keres2005}, \citealt{Dekel2006}, \citealt{Ocvirk2008}, \citealt{Keres2009}, \citealt{Brooks2009}, \citealt{vandeVoort2011}). The dense filaments have a short cooling time and can thus deliver cold gas to the galaxy where it can rapidly form stars (e.g., \citealt{Katz2003}, \citealt{Keres2005}, \citealt{FaucherGiguere2011}). At high redshifts ($z > 2$) halos with M$_{\rm halo} \lesssim 10^{12}$M$_{\odot}$ accrete primarily through the cold mode, while massive galaxies residing in halos with M$_{\rm halo} \gtrsim 10^{12}$ M$_{\odot}$ have their accretion dominated by the hot mode and they host a higher fraction of hot gas than cold gas in their halo \citep{Gabor2012}. Therefore, the fractional supply of cold halo gas available for future star-formation is lower in more massive galaxies. Simulations further show that the star-formation rate per unit mass is lower in these more massive systems \citep{Keres2012}. This accretion history naturally leads to less efficient star-formation and eventually a higher quenched fraction for galaxies with higher mass when the high fraction of hot gas is coupled with mechanisms, such as feedback processes, that prevent cooling.

At later epochs ($z \lesssim 2$), for galaxies residing in clusters, additional environmental quenching mechanisms such as ram pressure stripping, harassment, strangulation, and radio mode AGN feedback become important. Observational studies have found evidence of clusters with established intracluster media (ICM) as early as $z=2.5$ \citep{Wang2016} and $z=2.07$ \citep{Gobat2011}. In such environments, as a massive galaxy falls towards the center of the cluster, ram pressure stripping can strip cold gas from its outer disk (e.g., \citealt{Gunn1972}, \citealt{Giovanelli1983}, \citealt{Cayatte1990}, \citealt{Koopmann2004}, \citealt{Crowl2005}, \citealt{Singh2019}) if the pressure exerted by the ICM exceeds the restorative gravitational pressure provided by the galaxy. In the simplified treatment of \cite{Gunn1972}, this happens when:
\begin{eqnarray}
\rho_{\scalebox{.9}{$\scriptscriptstyle \rm ICM$}} V_{\rm infall}^2 > 2\pi \Sigma_\star \Sigma_{\rm gas}
\end{eqnarray}
where $\rho_{\scalebox{.9}{$\scriptscriptstyle \rm ICM$}}$ is the density of the ICM, $V_{\rm infall}$ is the component of the infalling galaxy's velocity perpendicular to its outer disk, and $\Sigma_\star$ and $\Sigma_{\rm gas}$ are the stellar and gas surface density in the disk of the infalling galaxy. Ram pressure stripping is particularly effective in stripping cold gas from the outer disk of galaxies where the gas and stellar surface densities are lower than in the central regions. This process is used to explain why the observed ratio of HI radius to optical radius is less than one for spiral galaxies in clusters while it is greater than one for field spirals (e.g.,  \citealt{Giovanelli1983}, \citealt{vanGorkom2011}), as well as the existence of truncated star-forming disks (e.g., \citealt{Cayatte1990}, \citealt{Koopmann2004}, \citealt{Crowl2005}). The effectiveness of ram pressure stripping depends on many factors including orientation, galaxy mass, gas density, and whether the stripped gas falls back to the disk and induces later star-formation (e.g., \citealt{Dressler1983}, \citealt{Gavazzi1993}, \citealt{Vollmer2001}, \citealt{Singh2019}).

In evolved clusters where galaxies have high velocity dispersions, harassment, the cumulative effect of many high speed tidal interactions (\citealt{Moore1996}, \citealt{Moore1998}) also becomes important. Harassment leads to gas inflows from the outer disk to the central regions via gravitational torques from induced bars and companions, causing high gas densities and high star-formation rates in the central regions of galaxies, at the expense of the outer disk.

Cluster galaxies can also be starved of fuel for future episodes of star-formation through strangulation, the slow removal of a cluster galaxy's hot gas reservoir through interactions with the cluster potential (e.g., tidal stripping) and ram pressure stripping by the cluster ICM (e.g., \citealt{Larson1980}, \citealt{Balogh2008}, \citealt{vandenBosch2008}). Starvation can lead to a slow decline of star-formation \citep{Larson1980}. Models that implement strangulation with delayed stripping (e.g., \citealt{Font2008}, \citealt{Cora2018}) rather than instantaneous stripping (e.g., \citealt{Springel2001}, \citealt{Kauffmann1993}) typically agree better with observational studies of the fraction of red and quiescent galaxies. An additional process that delays future star-formation in cluster galaxies is the radio mode of AGN feedback driven by powerful AGN jets (e.g., \citealt{Fabian2012}, \citealt{Heckman2014}) which heat surrounding cluster gas. This AGN feedback mode is observed directly through X-ray observations of the central galaxies of cool core clusters in the form of bubbles in the hot surrounding medium \citep{Fabian2012}.

In summary, a variety of mechanisms likely contribute to our empirical results which show that the quiescent fraction of massive galaxies increases as a function of stellar mass and that the quiescent fraction increases towards present day. Mergers, stellar and AGN feedback, and hot mode accretion play key roles across diverse environments, while mechanisms such as ram pressure stripping, harassment, strangulation, and radio mode AGN feedback become relevant at late times ($z \lesssim 2$) in cluster environments.

In order to observationally test the prevalence and impact of these quenching mechanisms, we are collecting and analyzing additional data. From a statistical perspective, the large comoving volume (0.33 Gpc$^3$ at $1.5 < z < 3.0$) probed by our study is expected to host a large number of massive dark matter halos and proto-clusters at $z < 3$. In order to identify potential proto-clusters and perform clustering analyses, spectroscopic redshifts are needed. These will be available in our field once the ongoing HETDEX optical spectroscopic survey is completed over the next several years (see Section \ref{sec:data}). In conjunction with the spectroscopic redshifts, our planned proposals to acquire deep high resolution space-based imaging will be able to reveal morphological signatures of tidal interactions and mergers (e.g., double nuclei, tails, arcs, ripples and other asymmetries), and will help to accurately estimate merger rates. We have also explored the relationship between AGN and star-formation activity in our field in \cite{Florez2020}. Deeper X-ray data will allow for a comprehensive study of the connection between AGN and star-formation, and their important contributions to massive galaxy evolution. 

\subsection{Comparing the Empirical Quiescent Fraction As a Function of Mass to Theoretical Predictions}
\label{sec:qf_mass_vTheory}

\begin{figure*}
\begin{center}
\includegraphics[width=\textwidth]{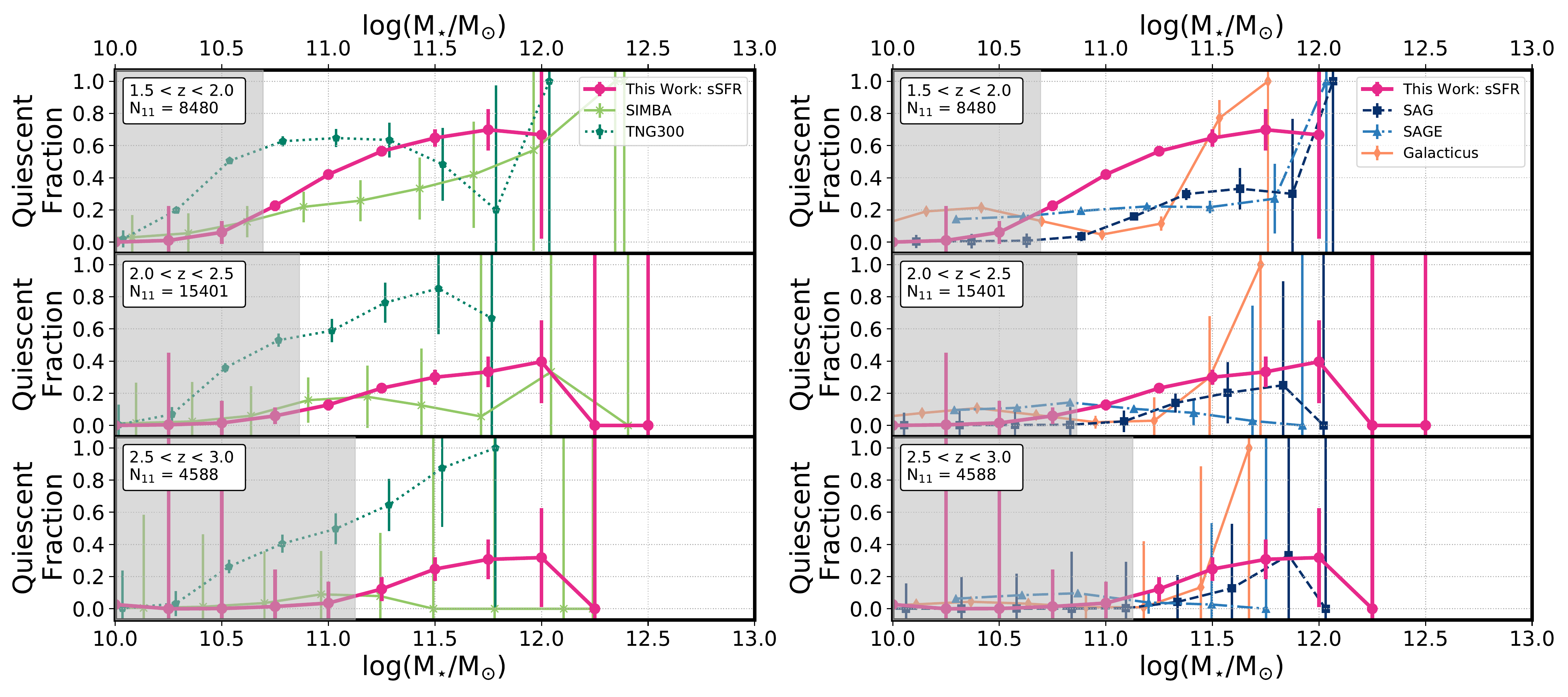} 
\caption{Empirical sSFR-selected quiescent fraction compared with sSFR-selected quiescent fractions from the hydrodynamical models SIMBA and IllustrisTNG (left) and semi-analytic models SAG, SAGE, and Galacticus (right). We find that results from the SIMBA model show a similar increase in quiescent fraction as a function of mass at $1.5 < z < 2.0$ as is seen with our empirical result, although the error bars are quite large. The results from IllustrisTNG show a quiescent fraction that increases as a function of mass at $2.0 < z < 3.0$, but decreases as a function of mass in the $1.5 < z < 2.0$ bin. We note that the quiescent fraction results from IllustrisTNG and any associated conclusions are highly dependent on the choice of aperture (see Appendix \ref{appendix:tng_aper}). SAM SAG is able to reproduce the trend seen in our empirical result as a function of mass, but underestimates the quiescent fraction by up to a factor of $\sim1.5-3$ compared with our empirical result. SAM SAGE does not predict a quiescent fraction that increases as a function of mass and it underestimates the quiescent fraction at the high mass end compared with our result, while Galacticus predicts a quiescent fraction that increases steeply at the high mass end, but with large error bars. The gray shaded region indicates masses below our completeness limit. Insets on the upper left of each panel show the number ($N_{11}$) of galaxies in our sample with $M_\star \ge 10^{11}$M$_\odot$.}
\label{qf_theory_ssfr_fig}
\end{center} 
\end{figure*} 

\begin{figure*}
\begin{center}
\includegraphics[width=\textwidth]{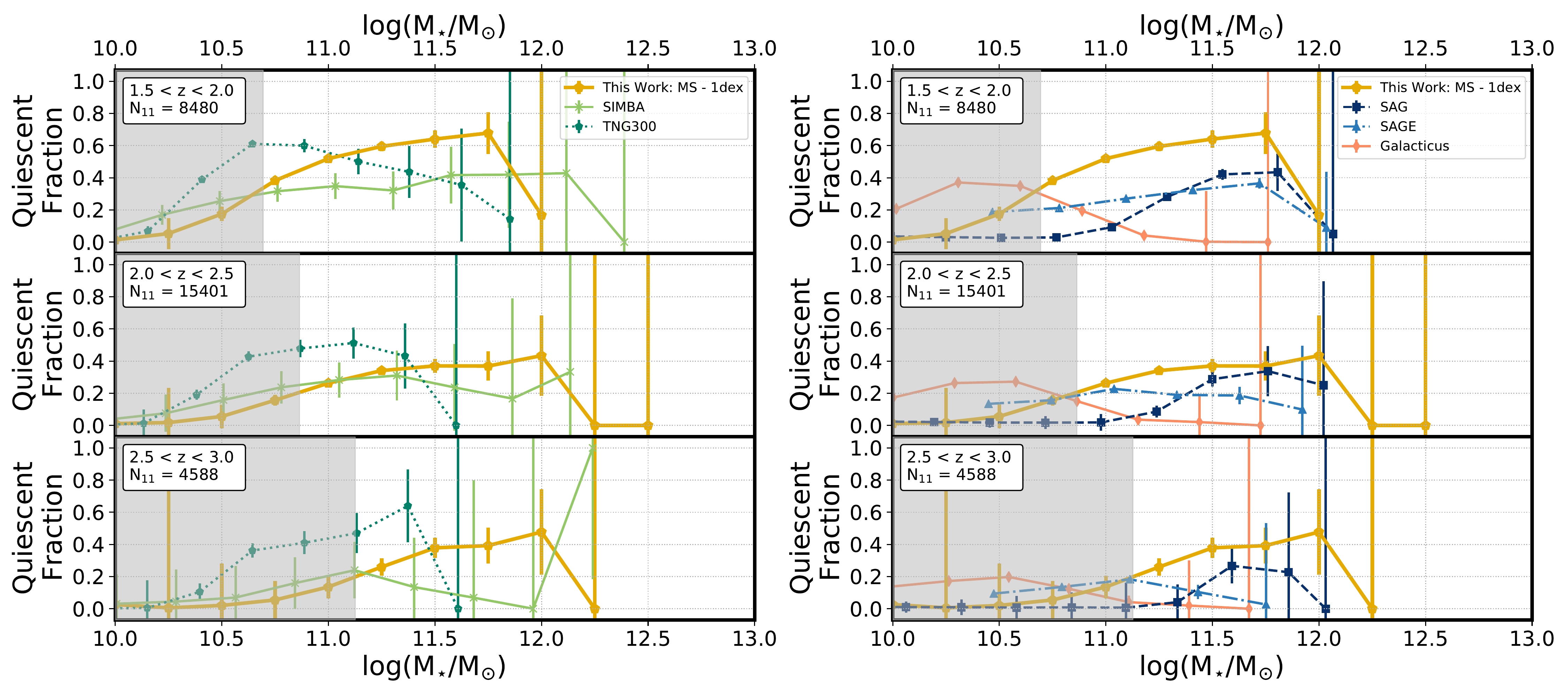} 
\caption{Empirical main sequence-selected quiescent fraction compared with the main sequence-selected quiescent fraction from hydrodynamical models IllustrisTNG and SIMBA (left) and semi-analytic models SAG, SAGE, and Galacticus (right). SIMBA under-predicts the quiescent fraction by a factor of $\sim1.5-4$ at $1.5 < z < 3.0$ compared with our empirical result, but with large error bars. The IllustrisTNG model over-predicts the quiescent fraction by a factor of $\sim2$ in the $2.0 < z < 3.0$ redshift bins compared with our empirical result and increases as a function of mass. In the lowest redshift bin ($1.5 < z < 2.0$), however, the IllustrisTNG quiescent fraction decreases as a function of mass. We note that the quiescent fraction results from IllustrisTNG and any associated conclusions are highly dependent on the choice of aperture (see Appendix \ref{appendix:tng_aper}). The three SAMS under-predict the quiescent fraction at the high mass end by up to a factor of 10 compared with our empirical result. SAG is the only SAM that predicts an increase in the quiescent fraction as a function of stellar mass. The gray shaded region indicates masses below our completeness limit. Insets on the upper left of each panel show the number ($N_{11}$) of galaxies in our sample with $M_\star \ge 10^{11}$M$_\odot$.}
\label{qf_theory_msdex_fig}
\end{center} 
\end{figure*}  

In Section \ref{sec:qf_shela} we showed that our empirical quiescent fraction increases as a function of mass in three redshift bins spanning $1.5 < z < 3.0$ and in Section \ref{sec:qf_shela_vz} we showed that the quiescent fraction for all massive galaxies ($M_\star \ge 10^{11}$M$_\odot$) increases as a function of redshift. Here, we compare our empirical results with two types of theoretical models: hydrodynamical models from IllustrisTNG (\citealt{Pillepich2018b}, \citealt{Springel2018}, \citealt{Nelson2018}, \citealt{Naiman2018}, \citealt{Marinacci2018}) and SIMBA \citep{Dave2019}, and semi-analytic models (SAMs) SAG \citep{Cora2018}, SAGE \citep{Croton2016}, and Galacticus \citep{Benson2012}. The goals of this comparison are to provide benchmarks to improve future implementations of theoretical models and to explore the implementations of physical processes (such as those discussed in Section \ref{sec:quench_mech}, including mergers, stellar and AGN feedback, hot mode accretion, ram pressure stripping, and tidal stripping) that drive massive galaxy evolution and impact the quiescent fractions predicted by these models.

For all of these models we compute the quiescent fraction for individual snapshots in the same way that we compute the quiescent fraction for our observed sample. These methods are described in detail in Section \ref{sec:qf_shela}. We focus our comparison with theoretical models on two of the three popular methods for measuring the quiescent fraction: sSFR-based and main sequence-based. For the sSFR-based method we use the same sSFR < $10^{-11}$ $\rm yr^{-1}$ threshold used for our data. To compute the main sequence-based quiescent fraction we define a main sequence for each model in the same way as is done for the data (where the main sequence is defined to be the average SFR in individual mass bins). We choose not to use the UVJ-based method for theoretical models because the results are strongly influenced by the chosen SEDs and dust laws used to extract photometry from the model. 

\subsubsection{Hydrodynamical Models}
\label{sec:qf_mass_vTheory_hydro}
IllustrisTNG is the latest generation of the Illustris hydrodynamical model that implements a variety of box sizes and improved feedback mechanisms. We utilize the largest volume available, $\sim$300$^3$Mpc$^3$ (TNG300; mass resolution $m_{\rm baryon}=1.1\times10^7~M_{\odot}$) with masses and star-formation rates measured within twice the stellar half mass radius (the $2 \times R_{1/2}$ aperture). We investigate how the choice of aperture impacts our comparison with IllustrisTNG quenched fractions in Appendix \ref{appendix:tng_aper}. When the SFR is below the IllustrisTNG resolution limit, the group catalogs have this value set to be SFR = 0. Following \cite{Donnari2019}, we reset this SFR to a random value between SFR = 10$^{-5}$ and 10$^{-4}$ before beginning our analysis to better represent the SFR for these objects. We note that the resulting quiescent fractions are the same whether we leave SFR = 0 or assign a random value between SFR = 10$^{-5}$ and 10$^{-4}$. 

The IllustrisTNG model implements both thermal-mode AGN feedback and kinetic-mode AGN feedback at high and low black hole accretion states, respectively (\citealt{Weinberger2017}, \citealt{Pillepich2018a}). The thermal-mode is a continuous injection of thermal energy into the gas surrounding the central black hole (heating the gas), while the kinetic-mode is a pulsed injection of kinetic energy to the regions near the black hole (acting as a wind). Stellar feedback is implemented through wind particles which are launched in random directions, where the strength of the wind is determined by the energy released from the supernova. 

We find that the IllustrisTNG model over-predicts the sSFR-based quiescent fraction by a factor of $\sim2-5$ and it over-predicts main sequence-based quiescent fractions by up to a factor of $\sim2$ at the high mass end ($M_\star \ge 10^{11}$M$_\odot$) compared to our empirical results (Fig. \ref{qf_theory_ssfr_fig} and Fig. \ref{qf_theory_msdex_fig}, respectively) at redshifts $1.5 < z < 3.0$. In the $2.0 < z < 3.0$ bins, the IllustrisTNG quiescent fractions increase as a function of mass, similarly to our empirical result. In the lowest redshift bin ($1.5 < z < 2.0$), however, the IllustrisTNG quiescent fraction turns over at $M_\star \simeq 10^{11}$M$_\odot$ and begins to decrease at the highest masses, albeit with large error bars. At log($M_{\star}$/$\rm M_{\odot}) \ge 11.5$ in the $1.5 < z < 2.0$ bin, IllustrisTNG under-predicts the quiescent fraction by up to a factor of $\sim3$ using both the sSFR- and main sequence-based methods compared with our empirical result. We note that the quiescent fraction results from IllustrisTNG and any associated conclusions are highly dependent on the choice of aperture, and we show the results using different aperture choices in Appendix \ref{appendix:tng_aper}. \cite{Donnari2019} also investigates the quenched fraction using the main sequence-based method and, using a different definition of the main sequence, they find a quiescent fraction that increases as a function of mass in all $1.5 < z < 3.0$ bins and is a factor of $\sim2-4$ larger than our empirical result. The comparison between these results is described in detail in Appendix \ref{appendix:tng_aper} and brings to light the importance of using the same methods when comparing quiescent fractions from different studies. 

We also compare our empirical result with that from the SIMBA cosmological simulation \citep{Dave2019} which implements meshless finite mass hydrodynamics in a 100 Mpc/h box with mass resolution $m_{\rm gas} = 1.82\times10^7~M_{\odot}$. We use the total stellar mass and SFR for galaxies in the SIMBA group catalog. The SIMBA model performs black hole feedback through both kinetic and X-ray modes. The kinetic-mode feedback is implemented as a bipolar wind (in the form of a collimated jet), and the X-ray mode injects energy into surrounding gas. We note that although SIMBA and IllustrisTNG both use the term ``kinetic-mode" to describe AGN feedback, their implementations of this feedback are quite different. Supernova feedback is implemented in SIMBA through a wind which carries hot and cold gas, as well as metals, away from star-forming regions. 

With the available SIMBA output (R. Dav{\'e}, private communication) we are able to compute the quiescent fraction using the sSFR and main sequence-based methods for their massive galaxies. We note that the small box size of the SIMBA cosmological simulation compared with the volume probed by our empirical sample leads to greater uncertainty at the highest masses. The SIMBA model tends to under-predict the quiescent fraction of massive galaxies when using both the sSFR-based method (Fig. \ref{qf_theory_ssfr_fig}) and the main sequence-based method (Fig. \ref{qf_theory_msdex_fig}) by a factor of $\sim1.5-4$ compared with our empirical result at $1.5 < z < 3.0$. We note that at $2.0 < z < 3.0$ the quiescent fraction predicted by SIMBA using both methods is largely flat as a function of mass, but in the lowest redshift bin ($1.5 < z < 2.0$), the SIMBA model begins to predict a quiescent fraction that increases as a function of mass. 

\begin{figure*}
\begin{center}
\includegraphics[width=\textwidth]{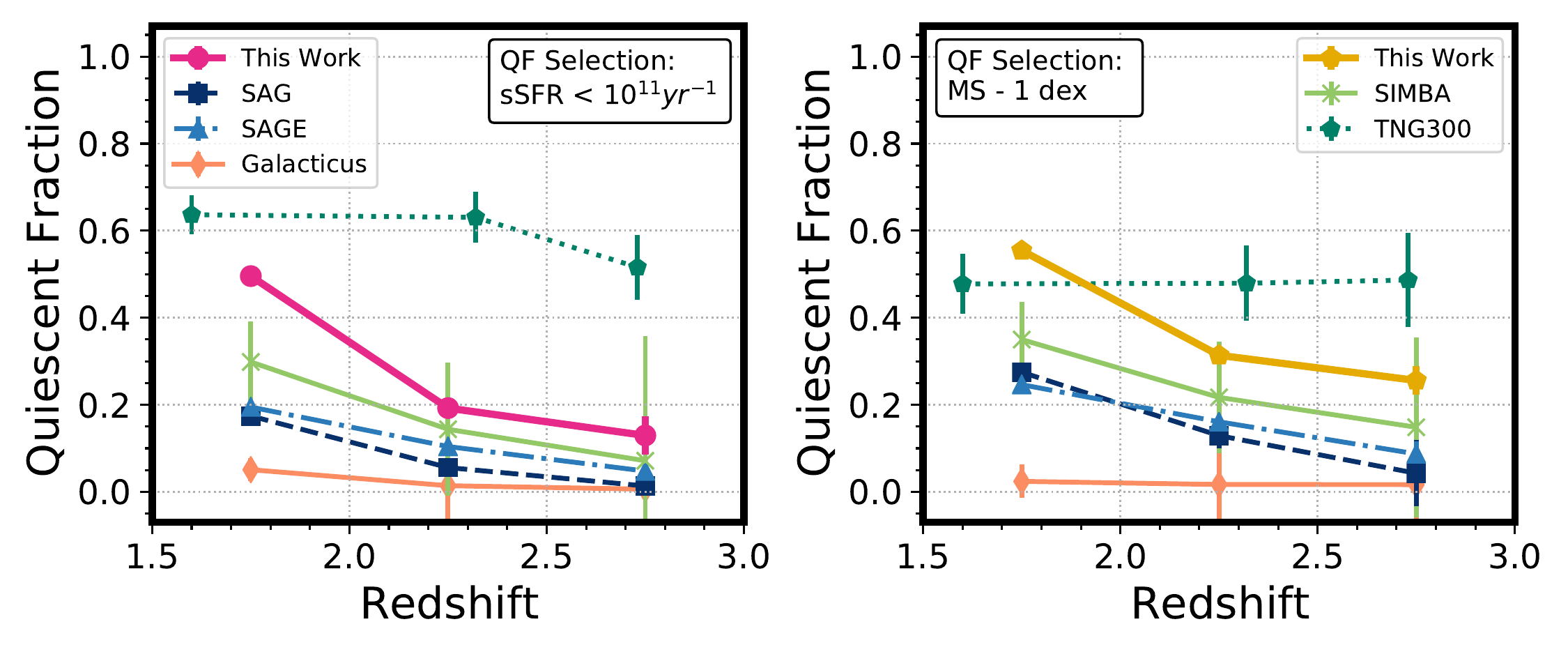} 
\caption{Empirical quiescent fraction for massive galaxies ($M_\star \ge 10^{11}$M$_\odot$) in our sample, selected using the sSFR (left) and main sequence (right) methods, compared with those from theoretical models. Theoretical models SIMBA, SAG, SAGE, and Galacticus under-predict the quiescent fraction for all massive galaxies by up to a factor of $\sim$10, while IllustrisTNG over-predicts the quiescent fraction by up to a factor of 3 compared with our empirical result. We note that results from IllustrisTNG and any associated conclusions are highly dependent on the choice of aperture (see Appendix \ref{appendix:tng_aper}). Using both the sSFR- and main sequence-based methods, all models except for Galacticus and IllustrisTNG predict a quiescent fraction that increases as a function of redshift. }
\label{qf_z_func_vTheory}
\end{center} 
\end{figure*} 

\subsubsection{Semi-Analytic Models}
\label{sec:qf_mass_vTheory_sams}
The results from three SAMs (SAG, SAGE, and Galacticus) are compared with our sSFR-based quiescent fraction (Fig. \ref{qf_theory_ssfr_fig}) and main sequence-based (Fig. \ref{qf_theory_msdex_fig}) empirical results. The benefit of using SAMs is that they are far less computationally expensive compared to hydrodynamical models and they provide a platform for exploring different analytical recipes for the physical processes underlying galaxy evolution. Additionally, SAMs efficiently model large volumes which allows for statistically significant samples of massive galaxies. 

The three SAMs we compare with, SAG \citep{Cora2018}, SAGE \citep{Croton2016}, and Galacticus \citep{Benson2012} are described in detail in the aforementioned works, as well as \cite{Knebe2018}, and will briefly be described here. The three SAMs implement their physical models to populate halos from the MultiDark-Planck2 (MDPL2) dark matter-only simulation, which has a box size of 1.0 $h^{-1}$Gpc on a side. The version of SAG used here (S. Cora, private communication) is run on 9.4\% of the full MDPL2 volume, while SAGE and Galacticus are run on the full 1.0 $h^{-1}$Gpc box. We use the total stellar mass and SFR for galaxies in the group catalogs for every SAM. Each of the SAMs implement different physical processes that influence galaxy evolution and the resulting quiescent fraction. We discuss the key points of these models (AGN and stellar feedback, treatment of mergers, the interaction of galaxies with the cluster potential, and the redshift at which the model is calibrated to observational results) here, but direct the reader to the original SAM papers for more detailed discussion. 

SAG implements AGN feedback through a radio-mode feedback scheme in which energy is injected into the region surrounding the black hole, reducing hot gas cooling. Stellar feedback heats gas within the galaxy and the energy transfer is regulated by a virial velocity and redshift dependence. The parameter that regulates the redshift dependence has been modified to generate the galaxy catalog used in this work (S. Cora, private communication). This parameter was adjusted to better reproduce the evolution of the star-formation rate density at high redshifts ($z>1.5$), and has also been shown to achieve a local quiescent fraction of galaxies that is in better agreement with observations from previous works \citep{Cora2018}. A fraction of the supernova ejecta is heated and removed from the halo. The ejected gas is reincorporated into the hot gas with a timescale that is inversely proportional to the corresponding (sub)halo virial mass. During major mergers, a starburst occurs in the bulge after stars and cold gas from the remnant are placed in the central regions. In a minor merger the stars of the less massive galaxy are transferred to the bulge of the more massive galaxy. A significant advantage of the SAG model is that it explicitly models ram pressure and tidal stripping for satellites falling into a group or cluster. Different stripping radii are used for the hot and cold gas components, as well as the disk and bulge stellar populations. The combined effects of ram pressure stripping (of gas) and tidal stripping (of stars and gas) are not instantaneous, rather the processes gradually remove the gas supply from a satellite. Calibration of SAG is performed considering observational constraints at $z=0$, $z=0.15$ and $z=2$. 

SAGE models AGN feedback through both radio- and quasar-modes. Radio-mode feedback heats the gas surrounding the black hole and keeps a history of past feedback events by implementing a hot gas region around the black hole which is not allowed to cool in subsequent time steps. This region of hot gas is only allowed to grow with time. The quasar-mode ejects cold gas (and hot gas if energetic enough) into a gas reservoir. Stellar winds from supernova feedback similarly eject gas into this reservoir. Gas in the reservoir is slowly incorporated back into the galaxy with the reincorporation regulated by the mass of the halo (higher mass halos receive more gas from the reservoir). In SAGE, a major merger results in the destruction of both disks and the stars are rearranged into a spheroid. Minor mergers simply move the satellite's gas and stars to the bulge of the central. SAGE implements a gradual stripping of satellites which is proportional to the stripping of the subhalo's dark matter, but it does not implement ram pressure stripping. The SAGE model is only calibrated to observational constraints at $z=0$. 

The Galacticus model implements AGN feedback through radio-mode which is an ejection of energy via a jet regulated by the black hole spin, and a quasar-mode wind. Stellar feedback is implemented as a wind that removes cold gas from the disk and into the hot halo. During major mergers, the gas and stars in merging galaxies are rearranged to form a spheroid, while in minor mergers the smaller galaxy is absorbed into the bulge of the massive galaxy. Ram pressure stripping and tidal stripping are not included in the model, however there is strangulation in which the hot atmosphere of satellites is stripped. Galacticus is only calibrated to observational constraints at $z=0$, and we note that the Galacticus model has not been calibrated to the MDPL2 dark matter simulation. 

We find that SAG does the best job of reproducing the trend of the sSFR-selected quiescent fraction that we find in our empirical results in all redshift bins where the quiescent fraction increases as a function of mass, however it under-predicts the quiescent fraction by up to a factor of $\sim 1.5-3$ at the high mass end compared to our empirical result. In contrast, SAGE finds that the sSFR-selected quiescent fraction is $\sim$0\% in the $2.5 < z < 3.0$ bin and in the $2.0 < z < 2.5$ bin it finds that the quiescent fraction decreases as a function of mass at the highest masses and under-predicts the quiescent fraction, compared with our empirical result, by a factor of $\sim2-5$. In the $1.5 < z < 2.0$ bin SAGE finds an sSFR-based quiescent fraction that is flat as a function of stellar mass and the quiescent fraction is lower than our empirical result by a factor of $\sim3$. Galacticus finds that the sSFR-selected quiescent fraction in the $2.0 < z < 3.0$ bins rises steeply as a function of mass, albeit with large error bars. In the $1.5 < z < 2.0$ bin, Galacticus under-predicts the quiescent fraction by up to a factor of 10 compared with our empirical result. 

The three SAMs underestimate the quiescent fraction of massive galaxies by up to a factor of $\sim$10 using the main sequence-based selection method in all of our redshift bins compared with our empirical result, however we note that SAG is able to correctly predict the increase in the quiescent fraction with increasing stellar mass. Galacticus predicts a quiescent fraction of $\sim$0\% at the high mass end and does not achieve a main sequence-based quiescent fraction that increases with increasing stellar mass. SAGE predicts a quiescent fraction that is flat in the $1.5 < z < 2.0$ bin and decreases at the high mass end in the $2.0 < z < 3.0$ bins. 

\subsection{Comparing the Empirical Quiescent Fraction As a Function of Redshift to Theoretical Predictions}
\label{sec:qf_z_vTheory}

\begin{figure*}
\begin{center}
\includegraphics[width=5.5in]{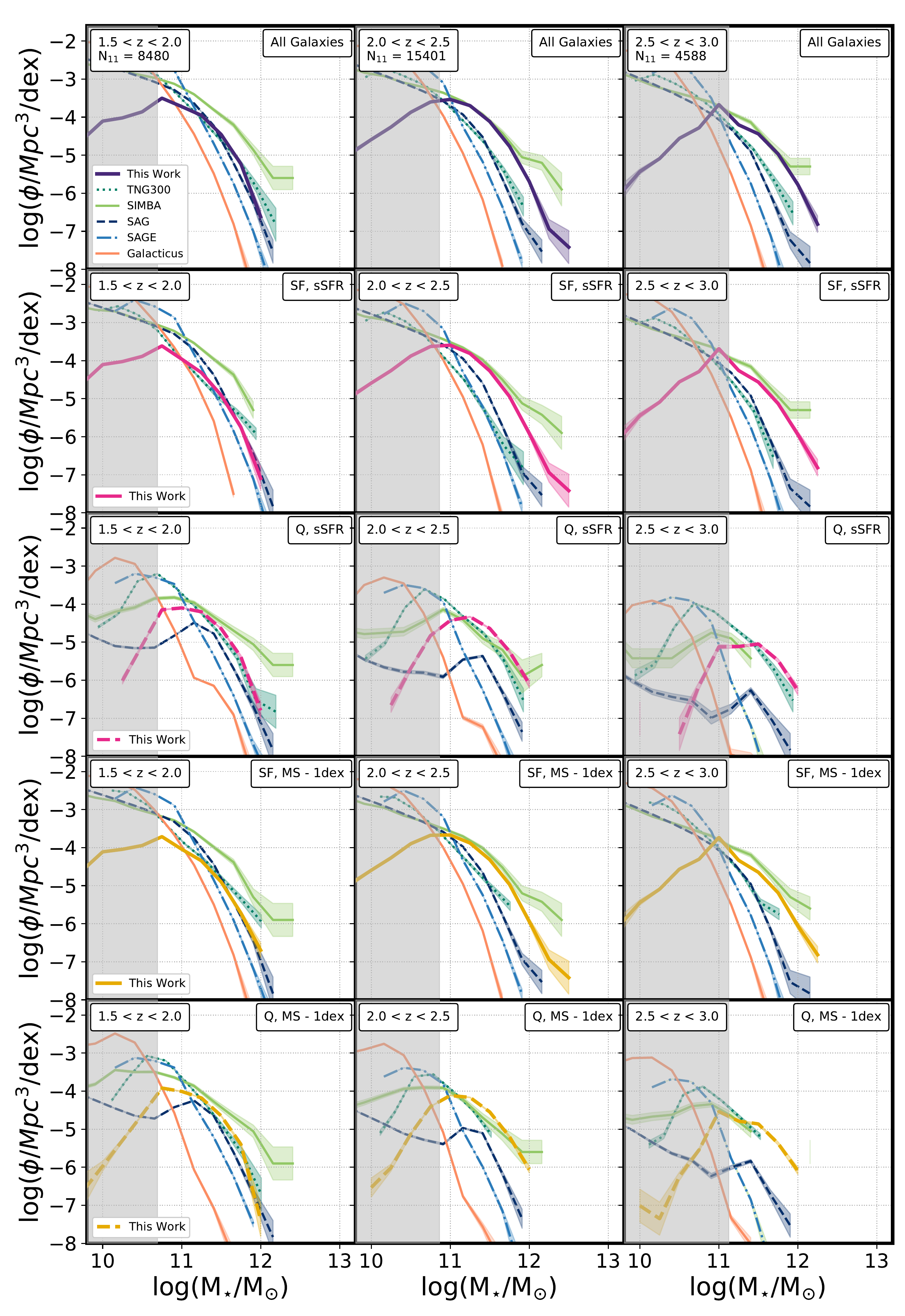} 
\caption{Empirical galaxy stellar mass function for our observed sample compared with predictions from hydrodynamical models SIMBA and IllustrisTNG and semi-analytic models SAG, SAGE, and Galacticus. The top row shows results for all galaxies. The 2nd and 3rd rows from the top show the star-forming and quiescent galaxy stellar mass functions, respectively, split into these populations using the sSFR-based method. The 4th and 5th rows from the top are analogous to rows 2 and 3, but split the quiescent and star-forming populations using the main sequence-based method. Poisson errors are indicated by the colored regions and are often smaller than the lines. The gray shaded region indicates masses below our completeness limit. In the top row, insets on the upper left of each panel show the total number ($N_{11}$) of galaxies in our sample with $M_\star \ge 10^{11}$M$_\odot$. A detailed comparison of our empirical results with those from theoretical models is given in Section \ref{sec:theory_smf}. Briefly, we find that hydrodynamical model SIMBA is in good agreement with our empirical total galaxy stellar mass function in our two redshift bins spanning $2.0 < z < 3.0$, while predictions from IllustrisTNG are lower than our result by a factor of 15 at these redshifts. We note that results from IllustrisTNG and any associated conclusions are highly dependent on the choice of aperture (see Appendix \ref{appendix:tng_aper}). The three SAMs under-predict the number density of the total population of massive galaxies by up to a factor of 10,000 compared with our empirical result, with the discrepancy being lower for SAG and SAGE than Galacticus. }
\label{smf_v_theory}
\end{center} 
\end{figure*} 

When exploring the quiescent fraction as a function of mass in individual redshift bins for theoretical models (Section \ref{sec:qf_mass_vTheory}), results are indicative of individual snapshots of the simulation. We can also compare the quiescent fraction for all galaxies with $M_\star \ge 10^{11}$M$_\odot$ as a function of redshift from theoretical models (Fig. \ref{qf_z_func_vTheory}) to our empirical result (see Section \ref{sec:qf_shela_vz}) to gain insights into the evolution of the massive galaxy population across several snapshots in the theoretical models.  

We find that hydrodynamical model SIMBA predicts quiescent fractions that increase from high- to low-redshift in a similar way to our empirical result for the sSFR-based method, however, SIMBA under-predicts the sSFR- and main sequence-based quiescent fractions for all galaxies with $M_\star \ge 10^{11}$M$_\odot$ by up to a factor of $\sim2$ compared with our empirical result. The IllustrisTNG quiescent fraction for all galaxies with $M_\star \ge 10^{11}$M$_\odot$ is largely flat as a function of redshift and over-predicts the sSFR- and main sequence-based quiescent fractions by factors of $\sim3$ and $\sim2$, respectively, compared with our empirical result. We note, again, that the quiescent fraction results from IllustrisTNG and any associated conclusions are highly dependent on the choice of aperture (see Appendix \ref{appendix:tng_aper}).

For both the sSFR-based and main sequence-based methods of identifying quenched galaxies, the three SAMs under-predict the quiescent fraction for all galaxies with $M_\star \ge 10^{11}$M$_\odot$ at $1.5 < z < 3.0$ compared with our empirical result, with the disagreement being larger for the main sequence-based method. Using the sSFR- and main sequence-based methods, semi-analytic model Galacticus under-predicts the quiescent fraction for all galaxies with $M_\star \ge 10^{11}$M$_\odot$ by up to a factor of $\sim$10 compared with our empirical result and is flat as a function of redshift. SAMs SAG and SAGE find quiescent fractions for all massive galaxies that increase as a function of redshift similarly to our empirical result using both the sSFR-based and main sequence-based methods. With the sSFR-based method, SAG and SAGE under-predict the quiescent fraction for all galaxies with $M_\star \ge 10^{11}$M$_\odot$ by a factor of $\sim3-4$ compared with our empirical result. Using the main sequence-based method, SAG and SAGE under-predict the quiescent fraction of massive galaxies by a factor of $\sim$5 compared with our empirical result. 

In order to better agree with our empirical result, theoretical models SIMBA, SAG, SAGE, and Galacticus might consider implementing physical processes that increase the fraction of quiescent galaxies in their massive galaxy population and reproduce the trend seen in our empirical result where the quiescent fraction increases as time progresses from $z=3.0$ to $z=1.5$. In contrast, IllustrisTNG may consider revisiting physical processes that decrease the quiescent fraction in their massive galaxy population in order to better agree with our empirical result. We note, again, that results from IllustrisTNG and any associated conclusions are highly dependent on the choice of aperture (see Appendix \ref{appendix:tng_aper}). SAM Galacticus and the hydrodynamical model from IllustrisTNG face particularly hard challenges as they both predict quiescent fractions which are flat as a function of redshift (Fig. \ref{qf_z_func_vTheory}). 

\subsection{Comparing the Stellar Mass Function to Theoretical Predictions}
\label{sec:theory_smf}

In Section \ref{sec:shela_smf} we showed the empirical galaxy stellar mass function for the total galaxy population, as well as the star-forming and quiescent galaxy populations. The total galaxy stellar mass function and quiescent galaxy stellar mass functions are connected through the quiescent fraction (see Eqn. \ref{eqn:smf}) and, therefore, the stellar mass function provides important insights when interpreting the quiescent fraction of massive galaxies. Comparisons of our empirical galaxy stellar mass functions and those predicted by theoretical models place important constraints on the physical process implemented in the theoretical models. In Section \ref{sec:shela_smf} we compared our result with those from previous observational studies and showed that in the two higher redshift bins ($2.0 < z < 2.5$ and $2.5 < z < 3.0$) our total stellar mass function is in fair agreement with previous studies, while in the lowest redshift bin ($1.5 < z < 2.0$), we may have a deficit of galaxies at masses log($M_{\star}$/$\rm M_{\odot}$) $< 11.2$. Therefore, when comparing our empirical galaxy stellar mass function to theoretical models, we treat this lowest redshift bin with caution and do not draw any strong conclusions from it.

The stellar mass functions from all theoretical models used in this work (hydrodynamical models from IllustrisTNG and SIMBA, SAMs SAG, SAGE, and Galacticus) are computed following the method of \cite{Tomczak2014} which is the same method used for our observed sample, but without a $1/V_{\rm max}$ correction as this is not necessary for theoretical models. Instead, the volume term is simply the volume of the simulation box for a given model. Additionally, we convolve the stellar mass functions from theoretical models with the average stellar mass error for our observed sample \citep{Kitzbichler2007}, which is computed from SED fitting (see Section \ref{sec:analysis}). 

We find that hydrodynamical model SIMBA is in agreement with the empirical number density of the total population of massive galaxies in the two redshift bins spanning $2.0 < z < 3.0$, while the prediction from IllustrisTNG is lower than our empirical result by up to a factor of 15 at these redshifts (top row Fig. \ref{smf_v_theory}). In section \ref{sec:qf_mass_vTheory_hydro} we showed that the SIMBA model under-predicts the quiescent fraction using both the sSFR and main-sequence methods by a factor of $\sim1.5-4$ at $1.5 < z < 3.0$ compared with our empirical result, and the IllustrisTNG model over-predicts the quiescent fraction using both methods by a factor of $\sim2-5$ compared with our empirical result. We note, however, that results from IllustrisTNG and any associated conclusions are highly dependent on the choice of aperture (see Appendix \ref{appendix:tng_aper}). In order to bring both the predicted quiescent fraction and stellar mass function into better agreement with our empirical result, the SIMBA hydrodynamical model may consider revisiting their implementation of processes that can increase the quiescent fraction across all three redshift bins spanning $1.5 < z < 3.0$, while maintaining the number density of the massive galaxy population at $2.0 < z < 3.0$. These processes include stellar and AGN feedback, ram pressure stripping, tidal stripping, strangulation, and harassment at $z<2.0$ when galaxy clusters are established. The IllustrisTNG model may consider implementing physical processes that will increase the number density of the massive galaxy population at early times ($2.0 < z < 3.0$). 

The SAMs SAG, SAGE, and Galacticus under-predict the number density of the total population of massive galaxies (top row Fig. \ref{smf_v_theory}) compared with our empirical result by up to a factor of $\sim$10,000 at the high mass end for our three redshift bins spanning $1.5 < z < 3.0$, with the discrepancy being smaller for SAG and SAGE than it is for Galacticus. The discrepancy decreases towards later epochs ($1.5 < z < 2.0$). In Section \ref{sec:qf_mass_vTheory_sams} we showed that the three SAMs under-predict the quiescent fraction of massive galaxies by a factor of $\sim1.5-10$, compared with our empirical result, using both the main sequence-based and sSFR-based quiescent galaxy selection. The SAMs show a larger level of disagreement with the total stellar mass function than they do with the quiescent fraction. In order to achieve better agreement with our empirical results, the three SAMs may consider improving the implementation of physical processes (see Section \ref{sec:quench_mech}) that can simultaneously alleviate the large disagreement with the empirical total galaxy stellar mass function and the moderate disagreement with the empirical quiescent fraction. Processes that could help to dramatically increase the total number density of massive galaxies include higher merger rates, different treatments of star-formation efficiency during mergers, and higher gas accretion. The models may be able to moderately increase their predicted quiescent fractions with the implementation of stronger ram pressure stripping, tidal stripping, strangulation, and stellar and AGN feedback.

\section{Summary}
\label{sec:summary}
Using multiwavelength data available in the 17.5 deg$^2$ SHELA footprint, we explore the buildup of the population of massive ($M_\star \ge 10^{11}$M$_\odot$) quiescent galaxies at cosmic noon as a function of stellar mass. We perform careful SED fitting to explore the growth and quenching of massive galaxies as a function of stellar mass at redshifts $1.5 < z < 3.0$. Our study benefits from the large area probed by our data which allows for small error bars dominated by Poisson statistics, as well as our uniform sample selection across large cosmic volumes giving an unbiased result. Our key results are summarized below. 
\begin{enumerate}
\item We implement three common techniques for measuring the quiescent fraction of massive galaxies: sSFR-based, main sequence-based, and UVJ-based selection techniques (Fig. \ref{qf_shela_plot}). Each of these three methods uses results from SED fitting to classify galaxies as either star-forming or quiescent. These three methods produce results that show an increase in the quiescent fraction as a function of mass and they are in good agreement at $1.5 < z < 2.0$ but differ by up to a factor of 2 at $2.0 < z < 3.0$.  As the stellar mass varies from log($M_{\star}$/$\rm M_{\odot}$) = 11 to 11.75 we find that at $2.5 < z < 3.0$ the main sequence-based quiescent fraction increases from $13.5\%\pm7.1\%$ to $39.6\%\pm11.2\%$, while at $1.5 < z < 2.0$ the quiescent fraction increases from $51.9\%\pm2.5\%$ to $66.4\%\pm13.1\%$. It is remarkable that by $z = 2$, only 3.3 Gyr after the Big Bang, the universe has quenched more than 25\% of massive ($M_\star = 10^{11}$M$_\odot$) galaxies. 
\\
\item We compare our empirical result using the UVJ-based quiescent fraction method with those from previous observational studies (Fig. \ref{qf_obs_uvj_fig}). Due to the larger volume probed by our 17.5 deg$^2$ study, our work at $1.5 < z < 3.0$ extends to higher stellar masses than earlier studies and our sample of galaxies with log($M_{\star}$/$\rm M_{\odot}$) $\ge$ 11.5 is a factor of $\sim$40 larger than samples from \cite{Muzzin2013}, \cite{Martis2016}, and \cite{Tomczak2016}. We find a similar trend with redshift (Fig. \ref{qf_z_func_vLit}), albeit with much smaller error bars and lower cosmic variance. 
\\
\item We explore several physical mechanisms that contribute to galaxy quenching across a range of environments and epochs (Section \ref{sec:quench_mech}). Additionally, we address which of these mechanisms can lead to our results that the quenched fraction increases as a function of mass at the high mass end and that the quenched fraction of massive galaxies increases towards present day. Across diverse environments, mergers, stellar and AGN feedback, and hot mode accretion play an important role, while in cluster environments mechanisms such as ram pressure stripping, harassment, strangulation, and radio mode AGN feedback likely become increasingly relevant for quenching star-formation.
\\
\item We also compare our empirical result with those from several classes of theoretical models (Fig. \ref{qf_theory_ssfr_fig} and Fig \ref{qf_theory_msdex_fig}). Hydrodynamical model IllustrisTNG over-predicts the main sequence-based quiescent fraction by a factor of 2 and the model over-predicts the sSFR-based quiescent fraction by a factor of 2 to 5 compared with our empirical result, however, the quiescent fraction results from IllustrisTNG and any associated conclusions are highly dependent on the choice of aperture (see Appendix \ref{appendix:tng_aper}). The SIMBA cosmological simulation tends to under-predict the quiescent fraction, compared with our empirical result, when using both the sSFR and main-sequence based methods by a factor of $\sim1.5-4$, respectively, but in the lowest redshift bin ($1.5 < z < 2.0$) it starts to correctly predict the observed trend of rising quiescent fraction with stellar mass. Semi-analytic models SAG, SAGE, and Galacticus tend to under-predict the quiescent fraction of massive galaxies using both the sSFR and main sequence-based selection techniques by a factor of $\sim1.5-10$ compared with our empirical result, however we note that SAG does the best job of recovering the trend of increasing quiescent fraction with increasing stellar mass found in our empirical result.  

Additionally, we compare these models to our empirical galaxy stellar mass function at the high-mass, steeply declining end (Fig. \ref{smf_v_theory}) for the total galaxy population, as well as the star-forming and quiescent populations. While the SAMs and the hydrodynamical model SIMBA under-predict the quiescent fraction of massive galaxies by a moderate factor of 1.5 to 10 at $1.5 < z < 3.0$ compared with our empirical result, we find that the SAMs drastically under-predict the number density of the total massive galaxy population by a factor of up to 10,000, while hydrodynamical models SIMBA and IllustrisTNG show only up to a factor of $\sim$15 disagreement (for IllustrisTNG, results and any associated conclusions are highly dependent on the choice of aperture; see Appendix \ref{appendix:tng_aper}). We discuss physical processes that might be revisited in the theoretical models to produce better agreement with the empirical quiescent fraction and stellar mass function.
\end{enumerate}

\vspace{5mm}
SS, SJ, and JF gratefully acknowledge support from the University of Texas at Austin, as well as NSF grant AST 1413652.  SS, SJ, JF, and SF acknowledge support from NSF grant AST 1614798. SS, SJ, JF, MS, and SF acknowledge generous support from The University of Texas at Austin McDonald Observatory and Department of Astronomy Board of Visitors. SS, SJ, JF, MS, and SF also acknowledge the Texas Advanced Computing Center (TACC) at The University of Texas at Austin for providing HPC resources that have contributed to the research results reported within this paper. SS is supported by the University of Texas at Austin Graduate Continuing Fellowship. LK and CP acknowledge support from the National Science Foundation through grant AST 1614668. SAC acknowledges funding from {\it Consejo Nacional de Investigaciones Cient\'{\i}ficas y T\'ecnicas} (CONICET; PIP-0387), and {\it Universidad Nacional de La Plata} (G11-150), Argentina. TH acknowledges CONICET, Argentina, for their supporting fellowships. CVM acknowledges financial support from the Max Planck Society through a Partner Group grant. The generation of SAG and SAGE data was performed on the OzSTAR national facility at Swinburne University of Technology. The OzSTAR program receives funding in part from the Astronomy National Collaborative Research Infrastructure Strategy (NCRIS) allocation provided by the Australian Government. SAC thanks Darren Croton for providing access to this facility. The authors wish to thank Annalisa Pillepich, Martina Donnari, and Mark Vogelsberger for providing IllustrisTNG results and useful comments, Romeel Dav{\'e} for providing SIMBA results and guidance, and Andrew Benson, and Darren Croton for their latest results and feedback regarding comparisons with semi-analytic models. The Institute for Gravitation and the Cosmos is supported by the Eberly College of Science and the Office of the Senior Vice President for Research at the Pennsylvania State University. This publication uses data generated via the Zooniverse.org platform, development of which is funded by generous support, including a Global Impact Award from Google, and by a grant from the Alfred P. Sloan Foundation.

\section*{Data Availability}
No new data were generated in support of this research.

\bibliographystyle{mnras}

\appendix

\renewcommand{\thefigure}{A\arabic{figure}}
\begin{figure*}
\begin{center}
\includegraphics[width=3in]{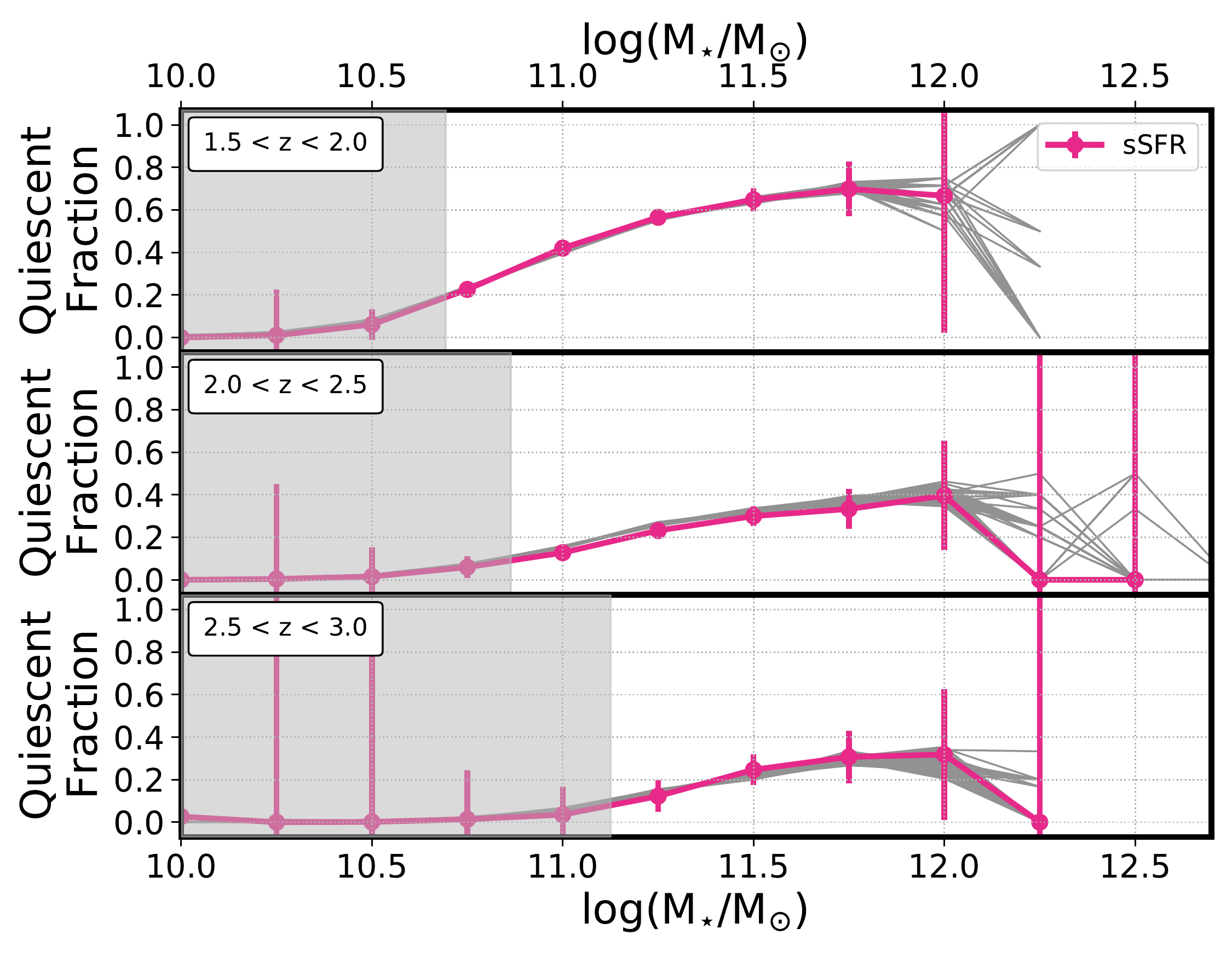} 
\includegraphics[width=3in]{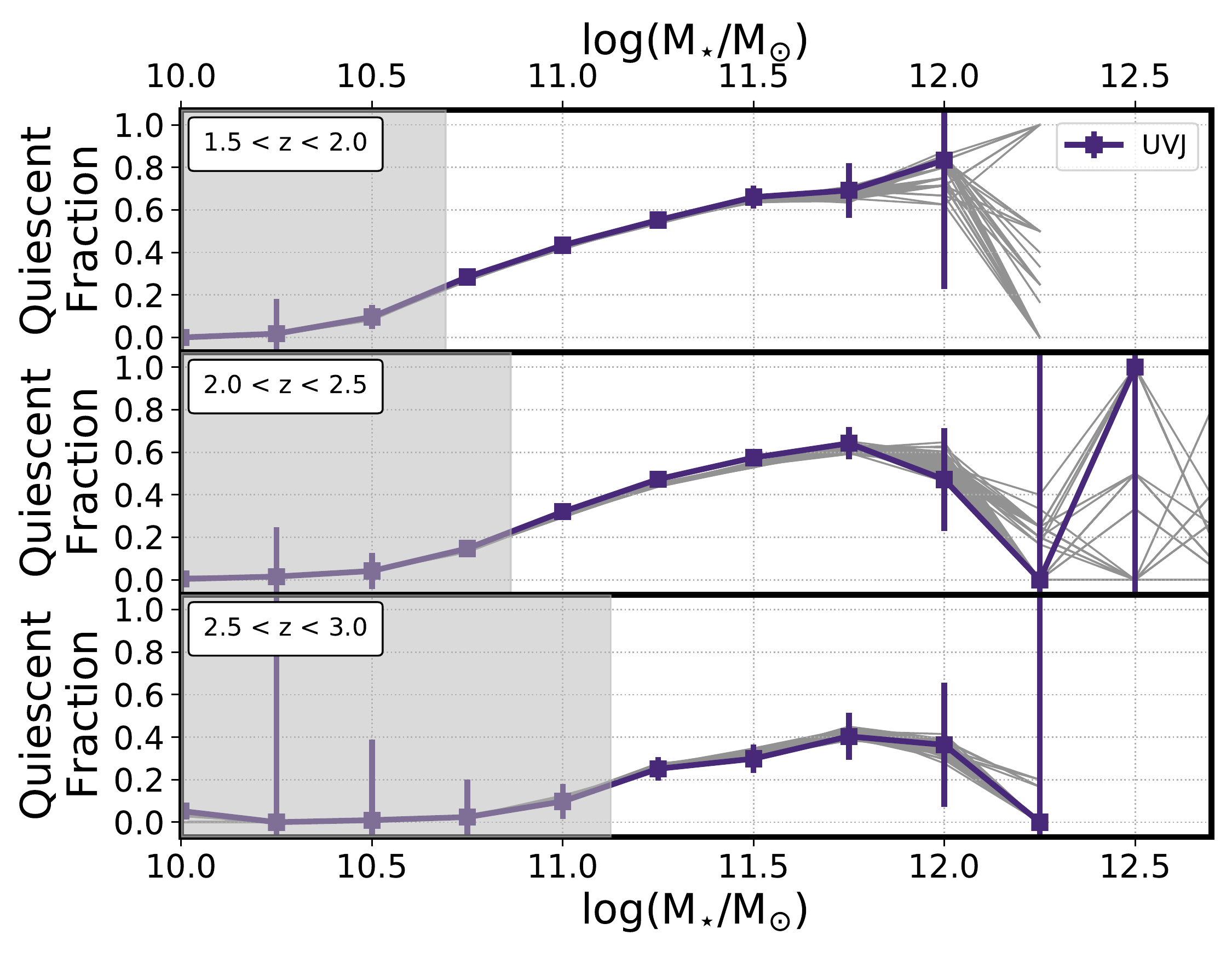} 
\includegraphics[width=3in]{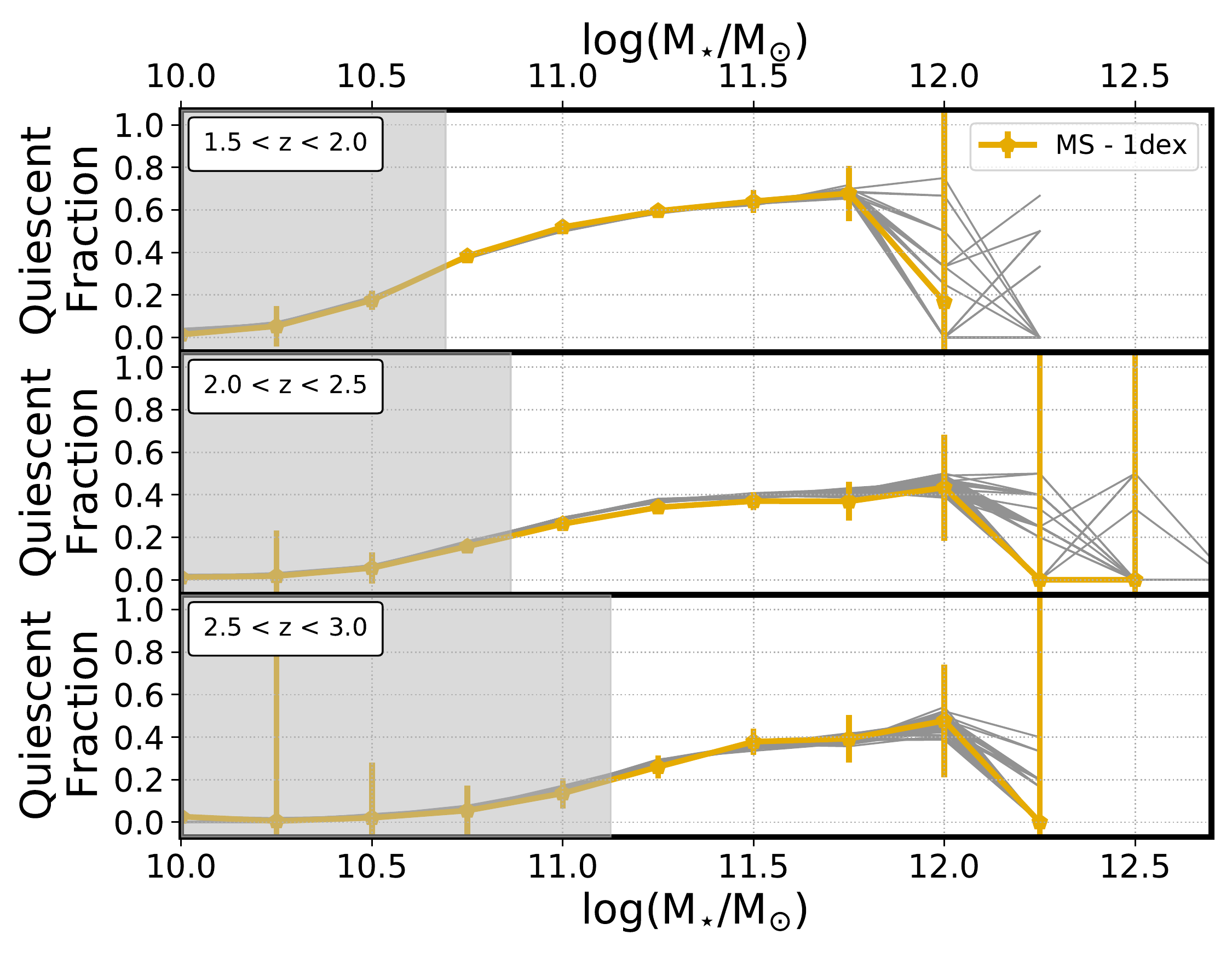} 
\caption{Results of the test in which we explore the impact of photometric redshift uncertainty and Eddington bias on our empirical quiescent fraction results. The three panels show the quiescent fraction measured using sSFR-selection (upper left), UVJ-selection (upper right), and distance from the main sequence selection (lower center). In each of the three panels, the quiescent fraction presented in Section \ref{sec:analysis} is shown as a colored line, and the results from the 100 catalog iterations are shown as grey lines. Our results are shown for each of the three redshift bins used throughout this work, which span $1.5 < z < 3.0$ and the redshift bin is indicated in the upper left of each row of the three panels. Using all three methods of measuring the quiescent fraction, the results of the 100 catalog iterations are consistent with the results presented in Section \ref{sec:analysis} for $M_\star = 10^{11} - 10^{12}$M$_\odot$, our mass range of interest throughout this work.}
\label{qf_iterCatalog}
\end{center} 
\end{figure*} 

\renewcommand{\thefigure}{A\arabic{figure}}
\begin{figure*}
\begin{center}
\includegraphics[width=\textwidth]{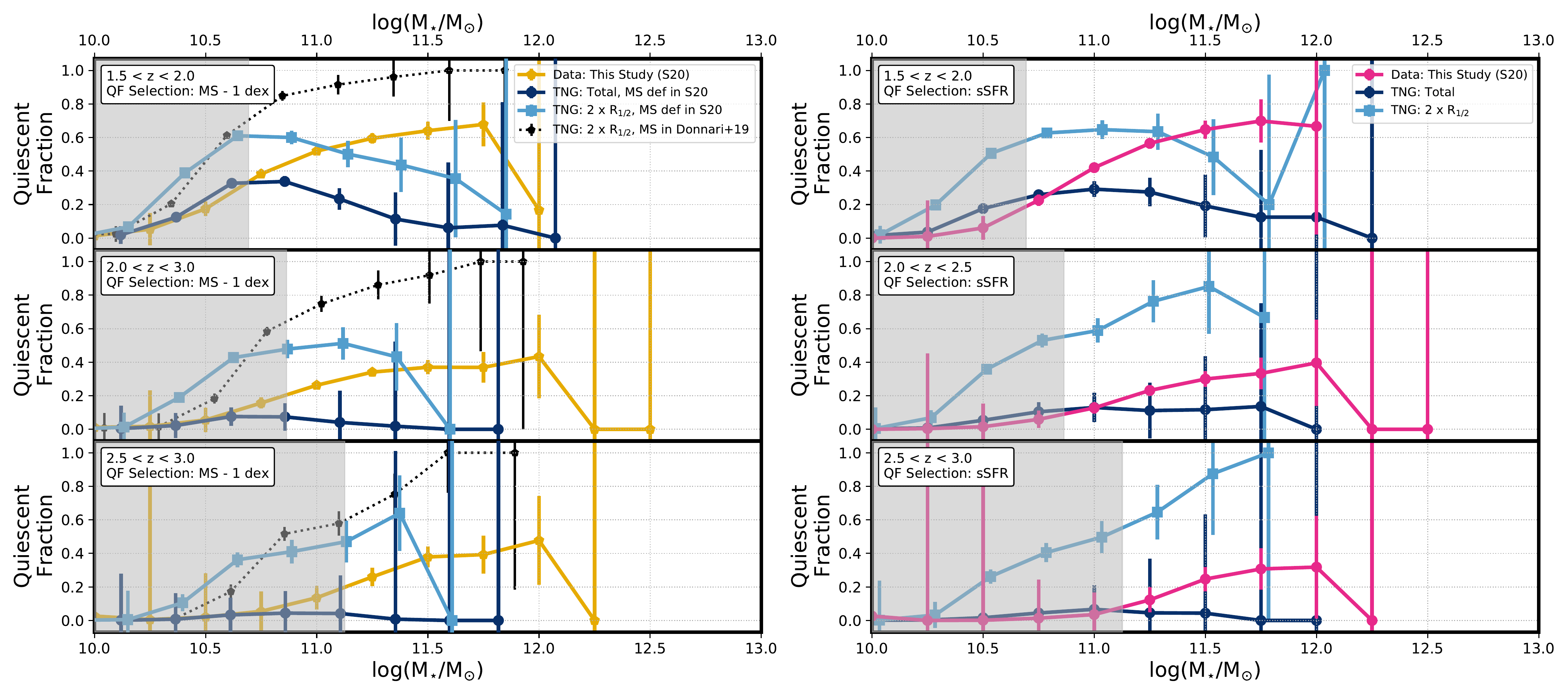} 
\caption{We compare the empirical quiescent fraction computed using the main sequence (left) and sSFR (right) methods with results from IllustrisTNG that are determined using different aperture types and main sequence definitions. Throughout this work we use the $2 \times R_{1/2}$ aperture (light blue) and the main sequence described in Section \ref{sec:qf_shela} to compute the quiescent fraction for the IllustrisTNG model and find that the IllustrisTNG main sequence- and sSFR-based quiescent fractions are higher than our empirical results. In contrast, when using masses and SFR measured for all particles bound to a subhalo (navy blue, the total ``aperture"), and the main sequence described in Section \ref{sec:qf_shela}, IllustrisTNG under-predicts the main sequence- and sSFR-based quiescent fractions compared to our empirical results. We also show the IllustrisTNG main sequence-based quiescent fraction from \protect\cite{Donnari2019} (black), which uses the $2 \times R_{1/2}$ aperture and a main sequence that is extrapolated from lower to higher masses, rather than a main sequence computed in every mass bin. This method gives a main sequence-based quiescent fraction that is higher than our empirical result.}
\label{qf_TNG_msdex_ssfr_apper}
\end{center} 
\end{figure*} 

\section{Testing the Impact of Photometric Redshift Error and Eddington Bias on the Empirical Quiescent Fraction}
\label{appendix:pertCatalog_test} 
In Section \ref{sec:analysis} we described a test in which we generate 100 iterations of our catalog by forcing our SED procedure to fit each galaxy in our catalog at 100 redshift values that are drawn from each galaxy's photometric redshift probability distribution. This test allows us to estimate the impact of photometric redshift uncertainty on our results and test the impact of Eddington bias \citep{Eddington1913}, the potential scattering of galaxies with low masses into a high mass bin. We find that our results from the 100 catalog iterations are consistent with our empirical results presented in Section \ref{sec:emperical_results} for $M_\star < 10^{12}$M$_\odot$. For $M_\star > 10^{12}$M$_\odot$ we find scatter in the quiescent fraction (up to a factor of $2-5$), which is expected at these extreme masses, and does not impact our results which focus on stellar masses $M_\star = 10^{11} - 10^{12}$M$_\odot$. The quiescent fraction for the 100 catalog iterations measured using the three methods employed throughout this work (sSFR, main sequence - 1 dex, and UVJ) are shown in Figure \ref{qf_iterCatalog}.

\section{Testing Different Aperture Choices and Main Sequence Definitions in Illustris TNG}
\label{appendix:tng_aper} 

Throughout this work we have computed the main sequence-based quiescent fraction using the main sequence definition described in Section \ref{sec:qf_shela}, where the main sequence is defined to be the average SFR in small mass bins and errors are computed using a bootstrap resampling procedure. We have also used the IllustrisTNG aperture where the stellar mass and SFR are measured within twice the stellar half mass radius (the $2 \times R_{1/2}$ aperture). Here, in Figure \ref{qf_TNG_msdex_ssfr_apper}, we show how the results differ if we use different aperture definitions or different definitions of the main sequence. 

We find that the IllustrisTNG main sequence-based (left panel Fig. \ref{qf_TNG_msdex_ssfr_apper}) and the sSFR-based (right panel Fig. \ref{qf_TNG_msdex_ssfr_apper}) quiescent fractions using masses and SFR measured within the $2 \times R_{1/2}$ aperture are over-predicted compared with our empirical result, as is shown throughout this work (see Section \ref{sec:qf_mass_vTheory_hydro}). If, instead, the total galaxy mass and SFR (computed for all star particles bound to a subhalo) are used, the IllustrisTNG quiescent fraction is lower than our empirical result by a factor of $\sim2-10$. Note that in making these comparisons with the main sequence-based quiescent fraction, we use the same main sequence definition used throughout this work and described in Section \ref{sec:qf_shela}. 

\cite{Donnari2019} investigated the quiescent fraction of galaxies in the IllustrisTNG model using the main sequence - 1 dex method out to $z=3$ (M. Donnari and A. Pillepich, private communication) using a different method to define the main sequence. Their method recursively computed a median SFR and removed galaxies 1 dex below that median until the median value (the main sequence) converged. This recursive procedure was done for all galaxies with stellar masses (measured using the $2\times R_{1/2}$ aperture) up to $M_\star \le 10^{10.2}$M$_\odot$ and was then linearly extrapolated to higher masses. With the extrapolated main sequence definition, \cite{Donnari2019} found a quiescent fraction that rises as a function of stellar mass and is a factor of $\sim2-4$ larger than our empirical quiescent fraction result in our three redshift bins spanning $1.5 < z < 3.0$ (left panel of Fig. \ref{qf_TNG_msdex_ssfr_apper}). The difference in quiescent fraction results that are computed using different aperture types and main sequence definitions highlight the extreme importance of comparing results achieved using the same methods. 

\bsp	
\label{lastpage}
\end{document}